\documentclass[12pt]{article}
\usepackage{epsfig}
\begin{document}
\begin{titlepage}
\begin{flushright}
  hep-th/0210025\\
  KUNS-1806\\
  HUPD-0209\\
\end{flushright}

\begin{center}
\vspace*{10mm}

{\LARGE \bf Wave-Function Profile and SUSY Breaking 
in 5D Model with Fayet-Iliopoulos Terms}
\vspace{12mm}

{\large
Hiroyuki~Abe$^{1,2,}$\footnote{
E-mail: abe@gauge.scphys.kyoto-u.ac.jp}, 
Tetsutaro~Higaki$^{1,}$\footnote{
E-mail: tetsu@gauge.scphys.kyoto-u.ac.jp} and 
Tatsuo~Kobayashi$^{1,}$\footnote{
E-mail: kobayash@gauge.scphys.kyoto-u.ac.jp}}
\vspace{6mm}

$^1${\it Department of Physics, Kyoto University,
Kyoto 606-8502, Japan}\\[1mm]
$^2${\it Department of Physics, Hiroshima University,
Hiroshima 739-8526, Japan}
\vspace*{15mm}

\begin{abstract}
We systematically study VEVs of a gauge scalar field $\Sigma$ in 
a bulk $U(1)$ vector multiplet and scalar fields in brane/bulk 
hypermultiplets charged under $U(1)$ in the 5D $S^1/Z_2$ 
orbifold model with generic FI terms. 
A non-trivial VEV of $\Sigma$ generates bulk mass terms for 
$U(1)$ charged fields, and their zero modes have non-trivial profiles. 
In particular, in the SUSY-breaking case, bosonic and 
fermionic zero modes have Gaussian profiles. 
Such non-trivial profiles are useful to explain 
hierarchical couplings. A toy model for SUSY breaking 
is studied, and it yields sizable $D$-term contributions 
to scalar masses. 
Because the overall magnitude of $D$-term contributions 
is the same everywhere in the bulk and also on both branes, 
we have to take into account these contributions and 
other SUSY-breaking terms to obtain a realistic description. 
We also give profiles and mass eigenvalues of higher modes. 
\end{abstract}

\end{center}
\end{titlepage}

\section{Introduction}
\label{sec:intro}
For the last several years, 4D $N=1$ 
supersymmetric models with an anomalous $U(1)$ gauge symmetry and 
non-vanishing Fayet-Iliopoulos (FI) $D$-term have been 
studied intensively, and many phenomenologically 
interesting aspects have been obtained.
Such anomalous $U(1)$ gauge symmetries, in general, appear 
in 4D string models.\cite{DSW,KN,typeI} 
This anomaly can be cancelled by the 4D Green-Schwarz mechanism.
For example, in weakly coupled 4D heterotic string models, 
the 4D dilaton field $S$ is required to 
transform non-trivially at the one-loop level as 
\begin{equation}
S \rightarrow S + {i \over 2 }\delta_{GS}\Lambda_X ,
\label{GS-S}
\end{equation}
under the $U(1)$ transformation with the transformation 
parameter $\Lambda_X$, where the Green-Schwarz coefficient 
is set as $\delta_{GS} = {\rm tr} Q_X /(48 \pi^2)$.
Then, a FI term $\xi$ is induced by the 
vacuum expectation value (VEV) of the dilaton field $S$,
\begin{equation}
\xi_S = {1 \over 2} \delta_{GS} \langle K'(S + \bar S)\rangle  M^2_P,
\label{GS-FI}
\end{equation}
where $M_P$ is the (4D) Planck scale, 
and $K'(S + \bar S)$ is the 
first derivative of the K\"ahler potential of the dilaton field.
At the tree level, we have $K(S + \bar S)=- \ln (S + \bar S)$,  
and the VEV of the dilaton field 
provides the 4D gauge coupling $g_4^2 = 1/\langle {\rm Re} S\rangle $. 
The FI term is suppressed by the one-loop factor
unless ${\rm tr} Q_X$ is quite large.
In type I models, other singlet fields, i.e. moduli fields, 
also contribute to the 4D Green-Schwarz mechanism, and a FI term is 
induced by their VEVs.\cite{typeI} 
Thus, the magnitude of the FI term is arbitrary in type I models, 
while its value in heterotic models 
is one-loop suppressed in comparison with $M_P$.

Some fields develop VEVs along $D$-flat directions 
to cancel the FI term, 
and the anomalous $U(1)$ gauge 
symmetry is broken around the energy scale $\xi^{1/2}$, 
e.g. just below $M_P$ in heterotic models.
Hence, this anomalous $U(1)$ gauge symmetry does not remain 
at low energy scales, although its discrete subsymmetry or 
the corresponding global symmetry may remain  even at 
low energy scales.
However, an anomalous $U(1)$ with non-vanishing FI term 
would create phenomenologically interesting effects.
Actually, several applications of anomalous $U(1)$ 
have been addressed.
One interesting application is the generation of hierarchical 
Yukawa couplings by the Froggatt-Nielsen mechanism.\cite{FN} 
We could use higher-dimensional couplings as the 
origin of hierarchical fermion masses and mixing angles, 
where the ratio $\xi^{1/2}/M_P$ plays a role in the derivation of 
the mass ratios and mixing angles like that of the 
Cabibbo angle.\cite{IR} 

Another aspect of anomalous $U(1)$ is that anomalous $U(1)$ 
symmetry breaking induces 
$D$-term contributions to soft supersymmetry (SUSY) breaking 
scalar masses.\cite{anomalous} 
These $D$-term contributions are, in general, 
proportional to the $U(1)$ charges, and 
their overall magnitude is on the order of other soft 
SUSY breaking scalar masses.
Thus, these $D$-term contributions become a new source 
of non-universal sfermion masses.
This would be dangerous from the viewpoint of 
flavor changing neutral currents (FCNCs). 
A type of SUSY breaking and the mechanism for its 
mediation by anomalous $U(1)$ and the 
FI term has been proposed.\cite{SUSY-breaking:mech} 
(See also Ref.~\cite{SUSY-breaking2}.)
In this type of scenario, $D$-term contributions 
can be much larger than other soft scalar masses and gaugino 
masses.
For example, in the case that only 
the top quark has vanishing $U(1)$ charge, 
stop and gaugino masses can be $O(100)$ GeV, and 
the other sfermion masses can be $O(10)$ TeV.
This type of sparticle spectrum can satisfy FCNC constraints 
as well as the naturalness problem, 
and actually, it corresponds to the decoupling solution 
for the SUSY flavor problem.\cite{decoupling} 
However, with this type of spectrum, we must be careful in 
regard to two-loop renormalization group effects on stop 
masses, which could become tachyonic.\cite{stop} 

Also, the FI term plays a role in the so-called 
$D$-term inflation scenario.\cite{cosmology,inflation,D-inf} 
Other interesting behavior has been studied 
in models with anomalous $U(1)$ and a nonvanishing FI term.

In addition to these 4D models, which were studied several 
years ago, recently 5D $S^1/Z_2$ models with the FI term 
were studied, and interesting behavior was 
found.\cite{Peskin}-\cite{Barbieri:2002ic} 
The $S^1/Z_2$ extra dimension with radius $R$ 
has two fixed points, which we denote $y=0$ and $\pi R$.
In general, we can put two independent FI terms on 
these two fixed points, i.e. $\xi_0$ and $\xi_\pi$.
In Ref.~\cite{Peskin}, SUSY breaking in the bulk is studied 
using the FI term.
In Refs.~\cite{AGW,KT,Pomarol}, zero-mode wave-function profiles 
are studied, and it is found that their profiles depend on the 
$U(1)$ charges. Hierarchical Yukawa couplings have also been 
derived.\cite{KT,Pomarol} 
In these models, the scalar field $\Sigma$ in the $U(1)$ 
vector multiplet (in the terminology of 5D $N=1$ SUSY ) 
plays an important role. 
In particular, bulk field profiles are studied 
systematically in Ref.~\cite{Nilles}, 
which considers models in which the sum of the 
FI coefficients vanishes, i.e. $\xi_0 + \xi_\pi = 0$.
This corresponds to a vanishing FI term 
in the 4D effective theory.

The interesting behavior exhibited by 4D models mentioned above 
can exist for non-vanishing FI terms.
Therefore, here we extend the analysis of Ref.~\cite{Nilles} 
to the case with $\xi_0 + \xi_\pi \neq 0$.
Actually, the SUSY breaking model of Ref.~\cite{Peskin} and 
the Yukawa hierarchy model of Ref.~\cite{KT} have 
FI terms for which $\xi_0 \neq 0$ and $\xi_\pi = 0$.
Thus, our analysis includes those cases.
We systematically study the VEV of $\Sigma$ 
in the generic case that $\xi_0$ and $\xi_\pi$ are independent. 
Then, we consider bulk field profiles.
Our result should be useful for further applications.

In this paper, we study the VEV of $\Sigma$ with 
generic values of $\xi_0$ and $\xi_\pi$, 
and we consider zero-mode wave-function profiles under the 
non-trivial VEV of $\Sigma$ in the bulk.
We also study mass eigenvalues and profiles 
of higher modes.
These systematic analyses are the main purpose of this paper.
In addition, we consider an application to SUSY breaking 
and examine the implication of $D$-term contributions 
to SUSY breaking scalar masses.

This paper is organized as follows.
In \S \ref{sec:2}, we systematically study the question of 
along which direction VEVs are developed in models with generic 
values of FI-terms.
Also, we consider the zero-mode profiles of bulk fields  
under non-trivial VEV of $\Sigma $ as well as 
profiles and mass eigenvalues of higher modes.
In \S \ref{sec:3}, as an example of an application, 
we introduce a toy model for SUSY breaking and 
discuss the implication of $D$-term contributions 
to SUSY-breaking scalar masses.
Section 4 is devoted to a conclusion and discussion. 
In the appendix, we investigate profiles and mass eigenvalues 
of higher modes.

\section{5D model with FI terms}
\label{sec:2}

\subsection{Setup}
\label{sec:2.1}

We consider a 5D SUSY model on $M^4 \times S^1/Z_2$.
Because of orbifolding, SUSY is broken into $N=1$ SUSY, 
in terminology of 4D theory.
Our model includes a $U(1)$ vector multiplet 
and charged hypermultiplets in the bulk. 
The 5D vector multiplet consists of a 5D vector field 
$A_M$, gaugino fields $\lambda_{\pm}$, a real scalar field 
$\Sigma$ and a triplet of auxiliary fields $D^a$.
Their $Z_2$ parities are assigned in a way consistent 
with 4D $N=1$ SUSY and $U(1)$ symmetry.\cite{Peskin} 
For example, the $Z_2$ parity of 
$\Sigma$ field as well as $A_5$ is assigned as odd. 
A 5D hypermultiplet consists of 
two complex scalar fields, $\Phi_{\pm}$, and their 
4D superpartners as well as their auxiliary fields.
The fields $\Phi_{\pm}$ have $Z_2$ parities $\pm$, and 
the $U(1)$ charge of the $Z_2$ even field $\Phi_+$ is 
denoted by $q$, while the corresponding $Z_2$ odd field 
has $U(1)$ charge $-q$.
$Z_2$ odd fields should vanish at the fixed points; e.g.,
\begin{equation}
\Sigma(0)=\Sigma(\pi R) = 0,\qquad 
\Phi_-(0) = \Phi_-(\pi R)=0.
\end{equation}
Also the first derivatives of $Z_2$ even fields 
along $y$ should vanish at the fixed points; e.g., 
\begin{equation}
\partial_y \Phi_+(0)= \partial_y \Phi_+(\pi R) =0.
\end{equation}
In addition to these bulk fields, 
our model includes brane fields, i.e. 
4D $N=1$ chiral multiplets on the two fixed points, 
and their scalar components are written 
$\phi_I$ ($I=0,\pi$), where $I$ denotes two 
fixed points.
Their $U(1)$ charges are denoted by $q_I$.
As a first step, we do not consider any 
superpotential on the fixed points.

We consider the following FI terms on the fixed points:
\begin{equation}
\xi (y) = \xi_0 \delta (y) +\xi_\pi \delta (y - \pi R).
\label{5D-FI}
\end{equation}
Here we simply assume these FI terms.
Our analysis is purely classical.
One-loop calculations are carried out in 
Ref.~\cite{Nilles3}-\cite{Barbieri:2002ic}, and they show that 
the FI terms are generated at the one-loop level.\footnote{
In Refs.~\cite{SSSZ,Nilles,Pomarol} it is shown that the 
one-loop generated FI terms include the second 
derivatives of delta-functions, $\delta''(y)$ and 
$\delta''(y-\pi R)$ as well as delta-function forms 
as in Eq.~(\ref{5D-FI}).
Here we do not consider such FI terms, but 
we can extend our analysis to the case with 
$\delta''(y)$ and $\delta''(y-\pi R)$.}
For one-loop generated FI terms, the sum of 
FI coefficients $\xi_0 + \xi_\pi$ is 
proportional to the sum of $U(1)$ charges.
Thus, in the case $\xi_0 + \xi_\pi \neq 0$, 
we have an anomaly, although such a case corresponds to 
the 4D models mentioned in the introduction, and we 
are interested in such a case.
It is not clear whether or not some singlet fields 
(brane/bulk moduli fields) transform non-trivially 
under the $U(1)$ transformation to cancel the anomaly and 
thereby generate FI terms like Eqs.~(\ref{GS-S}) and 
(\ref{GS-FI}).  Moreover, in what follows 
we discuss non-trivial profiles of bulk fields, 
and such profiles might change one-loop 
calculations of the FI term from the case with trivial 
profiles. 
Therefore, let us just assume the above FI terms 
and study their aspects classically.

Putting all the above consideration together, 
the scalar potential terms 
relevant to our study are written \cite{Peskin,AGW,Nilles}
\begin{eqnarray}
V &=& {1 \over 2}[-\partial_y \Sigma + \xi(y) +gq|\Phi_+|^2 - 
gq|\Phi_-|^2 + \sum_{I=0,\pi} gq_I|\phi_I|^2 
\delta (y-IR)]^2 \nonumber \\
&+& |\partial_y \Phi_+ -gq \Sigma \Phi_+|^2 
+  |\partial_y \Phi_- +gq \Sigma \Phi_-|^2 
+ g^2q^2|\Phi_+ \Phi_-|^2 +\cdots ,
\label{potential}
\end{eqnarray}
where $g$ is the 5D gauge coupling, which is related to $g_4$ 
as $g^2 = 2 \pi R g^2_4$.
We have taken a vanishing bulk mass, $m_\Phi =0$, for 
$\Phi_\pm$.
The first term corresponds to the $D$ term, and 
the other terms correspond to the $F$ terms.
Thus, the $D$-flat direction is written 
\begin{equation}
-\partial_y \Sigma + \xi(y) +gq|\Phi_+|^2 - 
gq|\Phi_-|^2+  \sum_{I=0,\pi} gq_I|\phi_I|^2 
\delta (y-IR) =0,
\label{D-flat}
\end{equation}
and the $F$-flat direction satisfies
\begin{eqnarray}
\partial_y \Phi_{\pm}  \mp 
gq \Sigma \Phi_{\pm} = 0,  \label{F-flat1} \\ 
\Phi_+ \Phi_- =0 .
\label{F-flat2}
\end{eqnarray}
It is convenient to integrate Eq.~(\ref{D-flat}), 
and the $D$-flat condition then becomes 
\begin{equation}
{1 \over 2}(\xi_0 +\xi_\pi) +\int^{\pi R}_0 dy 
(gq|\Phi_+|^2 - gq|\Phi_-|^2) +
{1 \over 2}  \sum_{I=0,\pi} gq_I|\phi_I|^2 =0.
\label{D-flat2}
\end{equation}
Here we have used the boundary conditions 
$\Sigma(0)=\Sigma(\pi R) =0$ for $\Sigma(y)$. 
Hereafter, we choose the signs of the $U(1)$ charges 
$q$ and $q_I$ such that $\xi_0+\xi_\pi \ge 0$.
We also denote the ratio of $\xi_\pi$ to $\xi_0$ as 
$r^\pi_0 = \xi_\pi / \xi_0$.

One of important aspect of the present system is that 
when $\Sigma$ develops a VEV, bulk masses of bulk 
matter fields are generated. When SUSY is not broken, 
the profiles of the zero modes of the $Z_2$ even fields $\Phi_+$ 
satisfy the equation 
\begin{equation}
\partial_y \Phi^{(0)}_{+} - gq\langle \Sigma\rangle \Phi^{(0)}_{+} =0; 
\label{phizeroee}
\end{equation}
that is, the profiles become 
\begin{equation}
\Phi^{(0)}_{+ }(y) =\Phi^{(0)}_{+ }(0)\exp 
\left[gq \int^y_0 dy' \langle \Sigma (y')\rangle \right]. 
\label{profile-1}
\end{equation}
Fermionic superpartners have the same profiles as the scalar 
fields (\ref{profile-1}).

If SUSY is broken and SUSY-breaking scalar masses 
are induced, the zero-mode profiles of the bulk scalar 
components would change, but the profiles of fermionic 
components do not change from their forms given in 
Eq.~(\ref{profile-1}). 
Furthermore, if SUSY-breaking scalar masses 
are constant in the bulk, zero-mode profiles of 
scalar components are not changed, but scalar mass eigenvalues  
are shifted from zero by SUSY-breaking scalar masses.
Thus, the VEV of $\Sigma$ is important.
In the following subsections, we study 
the VEV of $\Sigma$ and zero-mode profiles 
as well as higher modes systematically.

\subsection{The VEV of $\Sigma$ and the shape of zero modes}
\label{sec:2.2}

First, we consider the VEV of $Z_2$ odd bulk fields $\Phi_-$ 
in the SUSY vacuum.
The $F$-flat condition requires the VEV of $\Phi_-$ 
to satisfy the following equation: 
\begin{equation}
\langle \Phi_-(y)\rangle  =\langle \Phi_{-}(0)\rangle \exp 
\left[-gq \int^y_0 dy' \langle \Sigma (y')\rangle \right]. 
\label{vev-odd}
\end{equation}
However, the $Z_2$ odd bulk fields should satisfy 
the boundary condition $\Phi_{-}(0) =0$. 
This implies that $\langle \Phi_-(y)\rangle $ vanishes everywhere, 
unless $ \langle \Sigma\rangle $ has a singularity.
Thus, we consider vacua with $\langle \Phi_-(y)\rangle =0$ 
in the unbroken SUSY case.
Also, for the broken SUSY case studied in \S 2.2.1, we first consider 
the direction along $\langle \Phi_-(y)\rangle =0$, and then 
we discuss whether or not such a direction induces 
tachyonic masses for the first 
excited mode of $Z_2$ odd bulk fields.
Note that the zero modes of $Z_2$ odd bulk fields are 
projected out.
The first excited mode has a KK mass (e.g., of $O(1/R)$).
If the absolute value of negative $D$-term contributions 
to SUSY-breaking scalar (masses)$^2$ is larger than such KK 
(mass)$^2$, $Z_2$ odd bulk fields could also develop VEVs. 
However, we show that such first excited modes of $\Phi_-$ 
have non-tachyonic masses in the following section.
Thus, the vacuum with $\langle \Phi_-(y)\rangle =0$, 
which is studied in \S 2.2.1, corresponds to 
a local minimum.

\subsubsection{The case with $\langle \Sigma\rangle  \neq 0$ 
($^\forall q, q_I>0$)}
\label{sec:2.2.1}

First we consider the case that only $\Sigma$ develops a VEV.
This situation can be realized when 
$q > 0$ and $q_I > 0$ for the entire bulk and the brane fields 
except for $Z_2$ odd bulk fields. 

Varying $\Sigma$, we obtain the minimization condition,
$\partial^2_y \Sigma - \partial_y \xi(y) =0$.
Its solution is obtained as 
\begin{equation}
\partial_y \langle \Sigma \rangle = \xi(y) + C .
\label{vevphi}
\end{equation}
Since $\Sigma$ is a $Z_2$ odd field, it should satisfy 
the boundary conditions $\Sigma(0)=\Sigma(\pi R ) =0$. 
The integral constant $C$ is obtained from 
the following equation:
\begin{equation}
\int^{\pi R}_0 dy~\partial_y \langle \Sigma \rangle 
= {\xi_0 + \xi_\pi \over 2} +C\pi R.  
\end{equation}
The left-hand side of this equation should vanish, 
because of the boundary condition of $\Sigma$.
Thus we obtain
\begin{equation}
C =  -{\xi_0 + \xi_\pi \over 2\pi R} .
\label{valuec}
\end{equation}
The VEV $\langle \Sigma \rangle$ itself is written 
\begin{equation}
\langle \Sigma(y) \rangle = {1 \over 2 }\xi_0 {\rm sgn}(y) + Cy 
+{1 \over 2} \xi_\pi ({\rm sgn}(y-\pi R) +1) 
\label{VEVSigma-1}
\end{equation}
in the region $0 \leq y \leq \pi R$, 
where the function ${\rm sgn}(y)$ is defined as 
${\rm sgn}(y) = -1, 0,1$ for $y<0,y =0$ and $y>0$, 
respectively.
The case with $\xi_0 \neq 0$ and $\xi_\pi =0$ 
is discussed  in Ref.~\cite{Peskin}, 
and the case with $\xi_0+\xi_\pi =0$ is studied in 
Ref.~\cite{Nilles}. Thus, the above case includes those of 
these previous models.

Only with the above VEV of $\Sigma$, we have the vacuum energy 
\begin{equation}
V_4 = \int dy {1 \over 2}C^2= \int dy {1 \over 2}
\left( {\xi_0 + \xi_\pi \over 2 \pi R}\right)^2.
\end{equation}
Thus, in general, SUSY is broken.
However, in the case with $\xi_0 + \xi_\pi = 0$,  we 
have $C=0$, and SUSY is unbroken.
It is notable that even in the broken SUSY case, 
the vacuum energy is constant along the $y$ direction.

\begin{figure}[t]
\begin{center}
\begin{minipage}{0.48\linewidth}
   \centerline{\epsfig{figure=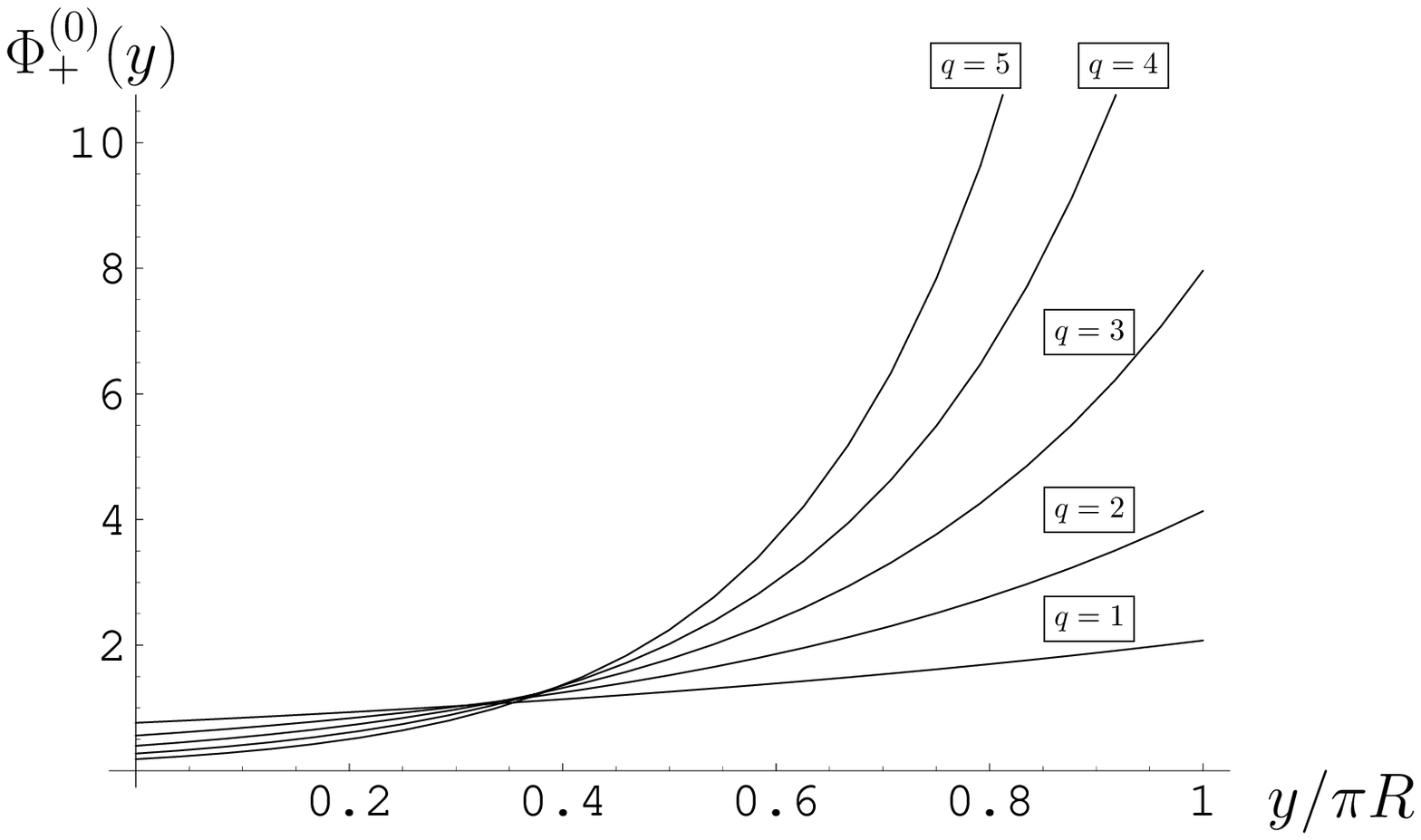,width=\linewidth}}
   \centerline{(a) $C=0$}
\end{minipage}
\end{center}
\begin{center}
\begin{minipage}{0.48\linewidth}
   \centerline{\epsfig{figure=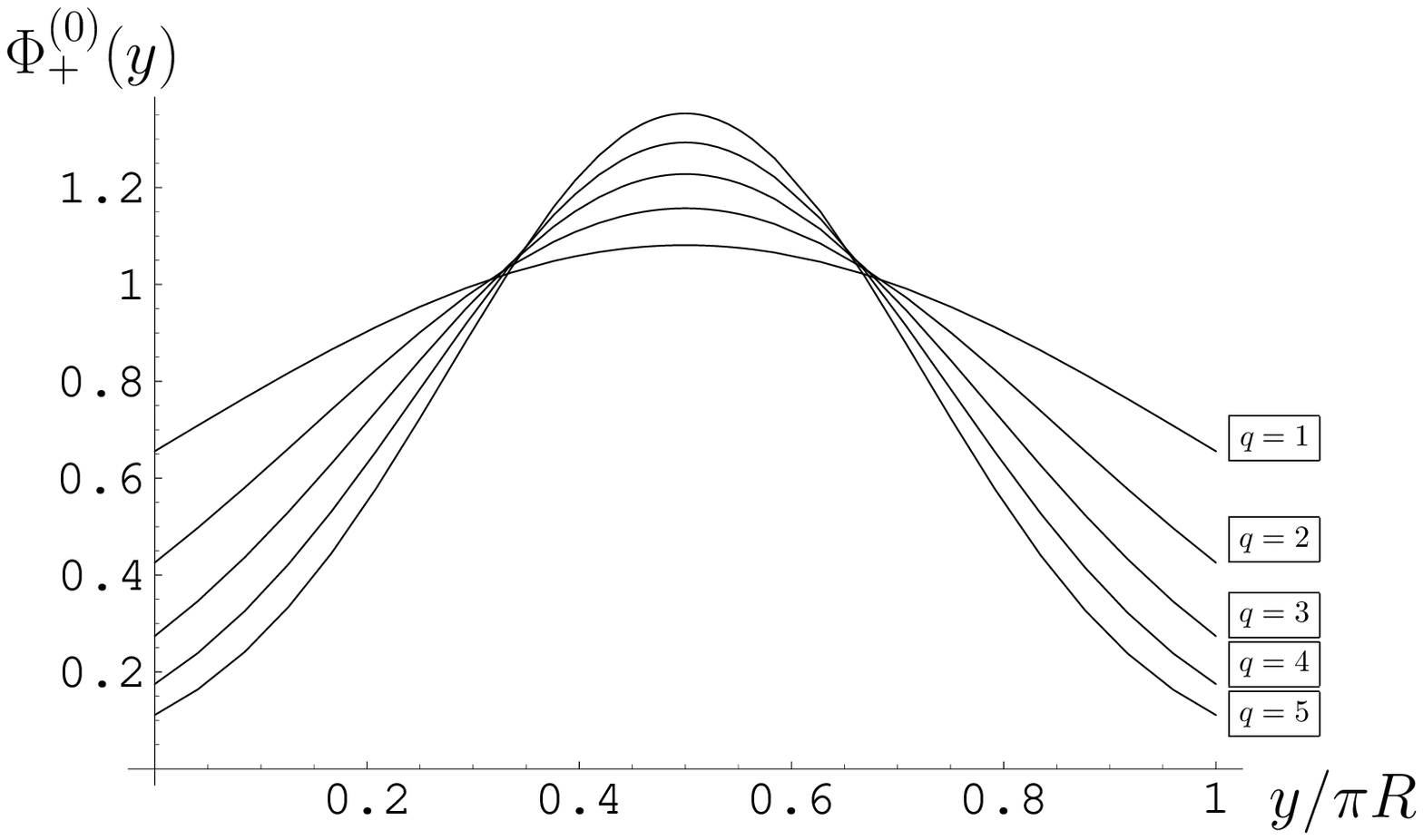,width=\linewidth}}
   \centerline{(b) $C \ne 0$ ($r^\pi_0=1$)}
\end{minipage}
\hfill
\begin{minipage}{0.48\linewidth}
   \centerline{\epsfig{figure=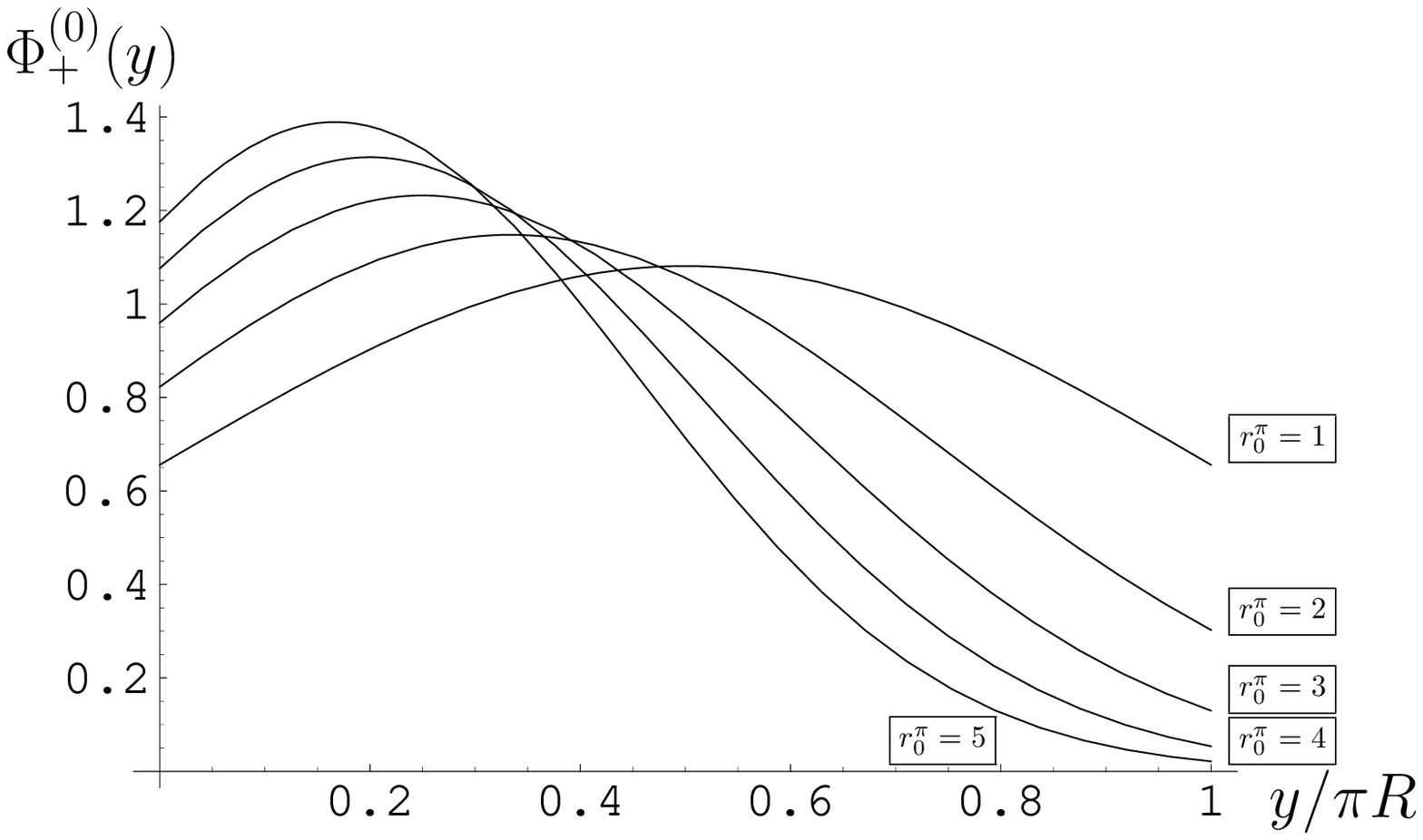,width=\linewidth}}
   \centerline{(c) $C \ne 0$ ($q=1$)}
\end{minipage}
\end{center}
\caption{Wave-function profile of the zero mode of 
$\Phi_+$. These are plotted by setting 
$g\pi R \xi_0/2 \equiv 1$ 
and varying $q$ or $r^\pi_0 = \xi_\pi/\xi_0$.}
\label{fig:gaussian}
\end{figure}
With the above VEV (\ref{VEVSigma-1}), we consider 
the profile of bulk fields.
As we show in Eq.~(\ref{Dterm-1}), even for the case 
$C \neq 0$, SUSY breaking scalar masses are constant along the 
$y$ direction. Thus, using Eq.~(\ref{profile-1}), we can write 
the profiles of both bosonic and fermionic  zero-modes of 
the $Z_2$ even bulk field as 
\begin{eqnarray}
\Phi_+(y) &=& A \exp[{gq \over 2}\xi_0 y] 
\qquad \qquad \qquad \qquad \qquad \qquad 
\ \ {\rm for} \ C=0, 
\label{profile-2}
\\
\Phi_+(y) &=& A \exp \left[-{gq \over 2\pi R}(\xi_0 +\xi_\pi)
\left(y-{\xi_0 \over \xi_0 + \xi_\pi}\pi R \right)^2 \right]
\quad {\rm for} \ C\neq 0 
\label{profile-3}
\end{eqnarray}
in the region $0 \leq y \leq \pi R$,
where $A$ is a suitable normalization factor. 
Note here, however, that scalar masses are shifted from zero by 
$D$-term contributions. 
The profile (\ref{profile-2}) is obtained in Refs.~\cite{AGW,Nilles}, 
and several aspects are discussed in Ref.~\cite{Nilles}.
The profile  (\ref{profile-3}) is a new result.
Actually this profile is interesting.
This form is a Gaussian profile localized at 
\begin{equation}
y = {\xi_0 \over \xi_0 + \xi_\pi}\pi R .
\end{equation}
These profiles are shown in Fig.~\ref{fig:gaussian}. 
Note that we are considering the case $q > 0$ 
for all $Z_2$ even bulk fields.
This localization point is independent of the $U(1)$ charges; 
that is, all of the zero modes are localized at the same point.
For example, if $\xi_0 = \xi_\pi$, all of the zero modes are 
localized at the point $y= (\pi R)/2$.
In some models the localization point may be outside the region 
$[0,\pi R]$, and in such cases, we see only the tail of 
the Gaussian profile.
Even in the limit of a very large FI term, i.e. $\xi_0, \xi_\pi \rightarrow 
\infty$, the ratio ${\xi_0 / (\xi_0 + \xi_\pi)}$ can be finite.
However, the width of the Gaussian profile 
depends on the $U(1)$ charges $q$.
This should be useful, e.g., to derive hierarchical couplings 
using wave-function overlapping.

If we include a non-vanishing bulk mass $m_\Phi$, 
the zero-modes become localized at different points. 
For example, the term $|\partial_y \Phi_+ -gq \Sigma \Phi_+|^2 $ 
in the scalar potential (\ref{potential}) changes as 
\begin{equation}
|\partial_y \Phi_+ -gq \Sigma \Phi_+|^2 
\rightarrow 
|\partial_y \Phi_+ +(m_\Phi - gq \Sigma )\Phi_+|^2 ,
\end{equation}
when we introduce a non-vanishing bulk mass $m_\Phi$.
This corresponds to a constant shift of 
$\langle \Sigma \rangle $ by $ m_\Phi/gq$.
Then, the localization points of the Gaussian profiles 
depend on the values of $m_\Phi$.
Hereafter, we set $m_\Phi =0$, but the following analysis and 
results can be simply extended to the case with non-vanishing 
$m_\Phi$ in a manner similar to that for the above constant shift of 
$\langle \Sigma \rangle $.

The scalar zero mode has a SUSY-breaking scalar mass term
due to the non-vanishing $D$ term in bulk
\begin{equation}
-gqC|\Phi_+|^2 ,
\end{equation}
which is constant along the $y$ direction, as stated above.
This scalar zero-mode mass is 
\begin{equation}
m^2_D = g q {\xi_0 + \xi_\pi \over 2 \pi R} .
\label{Dterm-1}
\end{equation}
Note that this $D$-term contribution is positive for the 
present case with $q>0$. 
Similarly, the brane fields on both branes have 
the $D$-term contribution to SUSY-breaking scalar mass  
\begin{equation}
m^2_{D,I} = g q_I {\xi_0 + \xi_\pi \over 2 \pi R}.
\label{Dterm-2}
\end{equation}
This $D$-term contribution is also positive 
in the present case with $q_I>0$.
There is no zero-mode with tachyonic mass.
Note that the overall magnitudes of $D$-term 
contributions to scalar masses are universal, up to 
$U(1)$ charges in bulk and both branes.
This point is considered again in \S \ref{sec:3}. 

Next, we discuss profiles and mass eigenvalues 
of the higher scalar modes $\Phi_{\pm}$.
{}From the potential (\ref{potential}), we have a $\Phi_\pm$ square 
term after $\Sigma$ develops a VEV $\langle \Sigma \rangle$. 
It is written 
\begin{eqnarray}
-{\cal L}_{\rm m}^{\Phi_\pm} &=& 
   \pm \left( \xi(y)-\partial_y \langle \Sigma \rangle \right) 
   gq |\Phi_\pm|^2 
 + |\partial_y \Phi_\pm \mp gq \langle \Sigma \rangle \Phi_\pm|^2
\nonumber \\ &=& 
-\Phi_\pm^\dagger \left( \partial_y^2 \mp gq \xi (y) 
 - (gq)^2 \langle \Sigma \rangle^2 \right) \Phi_\pm 
\equiv 
-\Phi_\pm^\dagger \Delta_\pm \Phi_\pm, 
\label{KKop}
\end{eqnarray}
where $\Delta_\pm \equiv \partial_y^2 \mp gq \xi(y) 
- (gq)^2 \langle \Sigma \rangle^2$. The eigenvalues of the operator 
$\Delta_\pm$ correspond to the KK spectrum of $\Phi_\pm$. 
Note that the term $\pm \left( \xi(y)-\partial_y \langle 
\Sigma \rangle \right) gq |\Phi_\pm|^2 $ is the SUSY-breaking 
scalar mass term.
Thus, we have to solve the equation 
\begin{equation}
\Delta_\pm \Phi_\pm + \lambda \Phi_\pm =0.
\end{equation}
In the appendix, its solutions are analyzed.
In this section, we give only results.

The case with $\xi_0 + \xi_\pi =0$, i.e. $C=0$, 
is discussed in Ref.~\cite{Nilles}.
(See also the appendix.)
Their (mass)$^2$ spectrum is obtained for 
$Z_2$ even and odd fields as 
\begin{equation}
m^2_k = {k^2 \over R^2} + {(gq\xi_0)^2 \over 4}. 
\qquad (k=1,2,\ldots) 
\label{czero}
\end{equation}

For the case with $C \neq 0$,  
the KK spectrum $\sqrt{\lambda_k}/M$ up to the 30th excited mode 
is shown in Fig.~\ref{fig:sp1} of the appendix, where 
$M$ is the representative scale of FI terms. 
{}From the figure, in the case $r^\pi_0=0$, we see that for 
$\Phi^{(k)}_\pm$, $m_k^2 \equiv \lambda_k$ behaves as 
\begin{eqnarray}
\begin{array}{llll}
\lambda_k & \sim & \left( \frac{M}{\pi} \right)^2 
\left\{ \left( \frac{\pi k}{RM} \right)^2 \pm \frac{1}{2} \right\}, & 
(k \gg RM) \\
\lambda_k & \sim & \left( \frac{M}{\pi} \right)^2 (2k\pm \frac{1}{2}), & 
(k \ll RM) 
\end{array}
\label{m-eigen1}
\end{eqnarray}
and in the case $r^\pi_0=1$, we obtain for $\Phi^{(k)}_\pm$ 
\begin{eqnarray}
\begin{array}{llll}
\lambda_k & \sim & \left( \frac{M}{\pi} \right)^2 
\left\{ \left( \frac{\pi k}{RM} \right)^2 \pm 1 \right\}, & (k \gg RM) \\
\lambda_k & \sim & \left( \frac{M}{\pi} \right)^2 (2k\pm 1), & (k \ll RM) 
\end{array}
\label{m-eigen2}
\end{eqnarray}
with the fixed parameters 
$\xi_0/\sqrt{\pi R} \equiv M^2/(\sqrt{2}\pi^2 g_4 q)$ in all cases. 
The case $k \ll RM$ is rather non-trivial, because $\lambda_k$ is 
linear in $k$ and it does not depend on $R$, the size of the extra dimension. 
This behavior can be understood as follows. Equation~(\ref{tPhiEEz}) 
is a Hermite differential equation (harmonic oscillator) 
for $R \to \infty$. Thus, for $k \ll RM$, we have a spectrum like 
that of a harmonic oscillator. 
On the other hand, when $\lambda$ is large compared with 
$g\xi_0/R$ and $gC$, the term with 
$\partial_y \tilde \Phi_{\pm}$ in Eq.~(\ref{tPhiEEz}) 
can be ignored. Thus, we have $\lambda \sim (k/R)^2$.
We also note that the first excited mode ($k=1$) of $\Phi_-$ 
does not become tachyonic, because $\lambda_k/M^2 \ge (2k-1)/\pi^2$ 
from Eqs.~(\ref{m-eigen1}) and (\ref{m-eigen2}).
Thus, the above vacuum is a local minimum.

The SUSY-breaking scalar mass terms are constants along the $y$ direction.
Hence, higher fermionic modes corresponding to  $\Phi_{\pm}$ have the 
same profiles as their scalar partners.
In Eqs.~(\ref{m-eigen1}) and (\ref{m-eigen2}), 
the terms $\pm M^2/(2\pi^2)$ and  $\pm M^2/(\pi^2)$ are 
the contributions from the SUSY breaking scalar masses. 
The corresponding fermionic modes do not have such contributions; 
that is, for example, for both $r^\pi_0 =0$ and $1$, we have 
\begin{eqnarray}
\begin{array}{llll}
\lambda_k & \sim & \left( \frac{M}{\pi} \right)^2 
 \left( \frac{\pi k}{RM} \right)^2, & (k \gg RM) \\
\lambda_k & \sim & \left( \frac{M}{\pi} \right)^2 2k. & (k \ll RM) 
\end{array}
\label{fm-eigen1}
\end{eqnarray}

In the case that $C$ is small compared with $\xi_0/R$, 
we can consider the term with $C$ in Eq.~(\ref{bulkEE}) 
to be a perturbation of the case with $C=0$. 
For example, the model discussed in \S \ref{sec:3} 
corresponds to such a case. 
At the first order of such a perturbation, (mass)$^2$ 
eigenvalues $\lambda_{k}$ are shifted from (\ref{czero}) as 
\begin{eqnarray}
\displaystyle
\begin{array}{ll}
\lambda_{k} = 
\displaystyle
  {k^2 \over R^2} + {(gq\xi_0)^2 \over 4} 
+ C \, 
\frac{\pi R(gq)^2 \xi_0}{e^{gq\xi_0 \pi R}-1} \times 
\frac{(k/R)^2+(gq\xi_0)^2/4}{(k/R)^2+5(gq\xi_0)^2/4}
& \textrm{for } \Phi_+, \\
\displaystyle
\lambda_{k} = 
  {k^2 \over R^2} + {(gq\xi_0)^2 \over 4} 
&  \textrm{for } \Phi_-.   
\end{array}
\end{eqnarray}
Note that for $\Phi_-$, the first-order perturbation in $C$ 
exactly cancels the $D$-term contribution (\ref{Dterm-1}).
There is no tachyonic mode for $\Phi_-$.
The mass eigenvalues of the corresponding fermionic modes are 
obtained by adding $\pm gqC$ for $\Phi_\pm$.

\subsubsection{The case with $\langle \Sigma \rangle, 
\langle \phi_I \rangle \neq 0$ ($^\exists q_I<0$)}
\label{sec:2.2.2}

To this point, we have considered the case 
$q > 0$ and $q_I > 0$ for all of the brane fields and $Z_2$ even 
bulk fields. In such case, only $\Sigma$ develops a VEV. 
Here, we consider another case, that in which there is a brane 
field with charge satisfying $q_I < 0$. 
In this case, such a brane field develops a VEV 
along the $D$-flat direction, and $U(1)$ is broken.
Our analysis does not depend on the brane on which 
the field with the charge $q_I < 0$ exists.
For concreteness, we choose this brane to be the $y=0$ brane.
The $D$-flat condition is satisfied with the VEV 
\begin{equation}
gq_0 \langle |\phi_0| \rangle ^2 = -(\xi_0 + \xi_\pi).
\end{equation}
Now, it is convenient to define the effective FI term,
\begin{equation}
\xi'(y) = \xi'_0 \delta (y) + \xi'_\pi \delta (y-\pi R),
\end{equation}
where $\xi'_0 = \xi_0 + gq_0\langle |\phi_0|^2 \rangle$ and 
$\xi'_\pi = \xi_\pi$. 
Note that for these effective FI coefficients, we have 
$\xi'_0 + \xi'_\pi =0$. 
For this case, the $D$-flat direction for $\langle \Sigma \rangle$ 
is obtained by replacing $\xi(y)$ by $\xi'(y)$ and setting 
$C=0$ in Eqs.(\ref{vevphi}) and (\ref{VEVSigma-1}), which yields 
\begin{equation}
\langle \Sigma(y) \rangle = {1 \over 2 }\xi'_0 {\rm sgn}(y) 
+{1 \over 2} \xi'_\pi ({\rm sgn}(y-\pi R) +1) .
\label{VEVSigma-2}
\end{equation}
Since we assume no superpotential for $\phi_0$, 
SUSY is unbroken along this direction.

The zero-mode profile of the $Z_2$ even bulk field is 
obtained by replacing $\xi(y)$ by $\xi'(y)$ in 
Eq.~(\ref{profile-2}), yielding 
\begin{equation}
\Phi_+^{(0)}(y) 
= A \exp \left[{gq \over 2}\xi'_0 y \right]. 
\end{equation}
This is effectively the same as the case studied 
systematically in Ref.~\cite{Nilles}.

Mass eigenvalues and wave-function profiles 
of higher modes are the same as those in the case 
$C=0$ considered in \S \ref{sec:2.2.1}, except that $\xi$ is replaced 
by $\xi'$; that is, (mass)$^2$ eigenvalues are obtained as 
\begin{equation}
m^2_k = {k^2 \over R^2} + {(gq\xi'_0)^2 \over 4}.
\end{equation}

\subsubsection{The case with $\langle \Sigma \rangle, \langle 
\Phi_+ \rangle \neq 0$ ($^\exists q<0$)}
\label{sec:2.2.3}

If there is a $Z_2$ even bulk field whose charge satisfies 
$q < 0$, it can also develop a VEV along the flat direction.
Here, we consider the case that $\Sigma$ and the bulk field 
$\Phi_+$ with charge  $q < 0$ develop VEVs.
In this case, the $D$-flat and $F$-flat conditions 
are written 
\begin{eqnarray}
-\partial_y \langle \Sigma \rangle + \xi (y) 
+gq \langle |\Phi_+| \rangle ^2 = 0, 
\label{D-flat3}
\\
\partial_y \langle \Phi_+ \rangle -gq 
\langle \Sigma \rangle \langle \Phi_+ \rangle =0 .
\label{F-flat3}
\end{eqnarray}
We multiply the former and latter equations here by 
$\langle \Sigma \rangle$ and $\langle \Phi^*_+ \rangle $, 
respectively, and compare them. We thereby obtain 
\begin{equation}
\partial_y(\langle \Sigma \rangle ^2 - 
\langle |\Phi_+ | \rangle ^2 ) = 2 \langle \Sigma 
\rangle  \xi(y).
\end{equation}
Integrating this equation and substituting 
the result into Eq.~(\ref{F-flat3}), 
we find that the following form of $\langle |\Phi_+ |\rangle$ 
is the solution: 
\begin{equation}
\langle |\Phi_+ (y)| \rangle = 
{A \over \cos [ay + b + c_0 {\rm sgn}(y) 
+ c_\pi {\rm sgn }(y-\pi R)]}.
\label{vevphi-3}
\end{equation}
Moreover, substituting this form of 
$\langle |\Phi_+ |\rangle$ into Eq.~(\ref{F-flat3}), 
we obtain 
\begin{equation}
\langle \Sigma (y) \rangle = {a \over gq}
\tan[ay + b+ c_0 {\rm sgn}(y) 
+ c_\pi {\rm sgn }(y-\pi R)].
\label{vevSigma-3}
\end{equation}
Actually, the solution constituted by (\ref{vevphi-3}) 
and (\ref{vevSigma-3}) is obtained 
in Ref.~\cite{KT} for the case $\xi_0 \neq 0$ 
and $\xi_\pi=0$. The flat direction represented by 
(\ref{vevphi-3}) and (\ref{vevSigma-3})
is the simple extension of that result to the case with 
generic values of $\xi_0$ and $\xi_\pi$.
Along this direction, SUSY is unbroken, but $U(1)$ is broken.
Here, Eq.~(\ref{D-flat3}) requires the constants to satisfy 
\begin{equation}
a^2 = g^2 q^2 A^2, \qquad {2a \over gq} c_I = \xi_I.
\label{sol-2}
\end{equation}
\begin{figure}[t]
\begin{center}
\begin{minipage}{0.48\linewidth}
   \centerline{\epsfig{figure=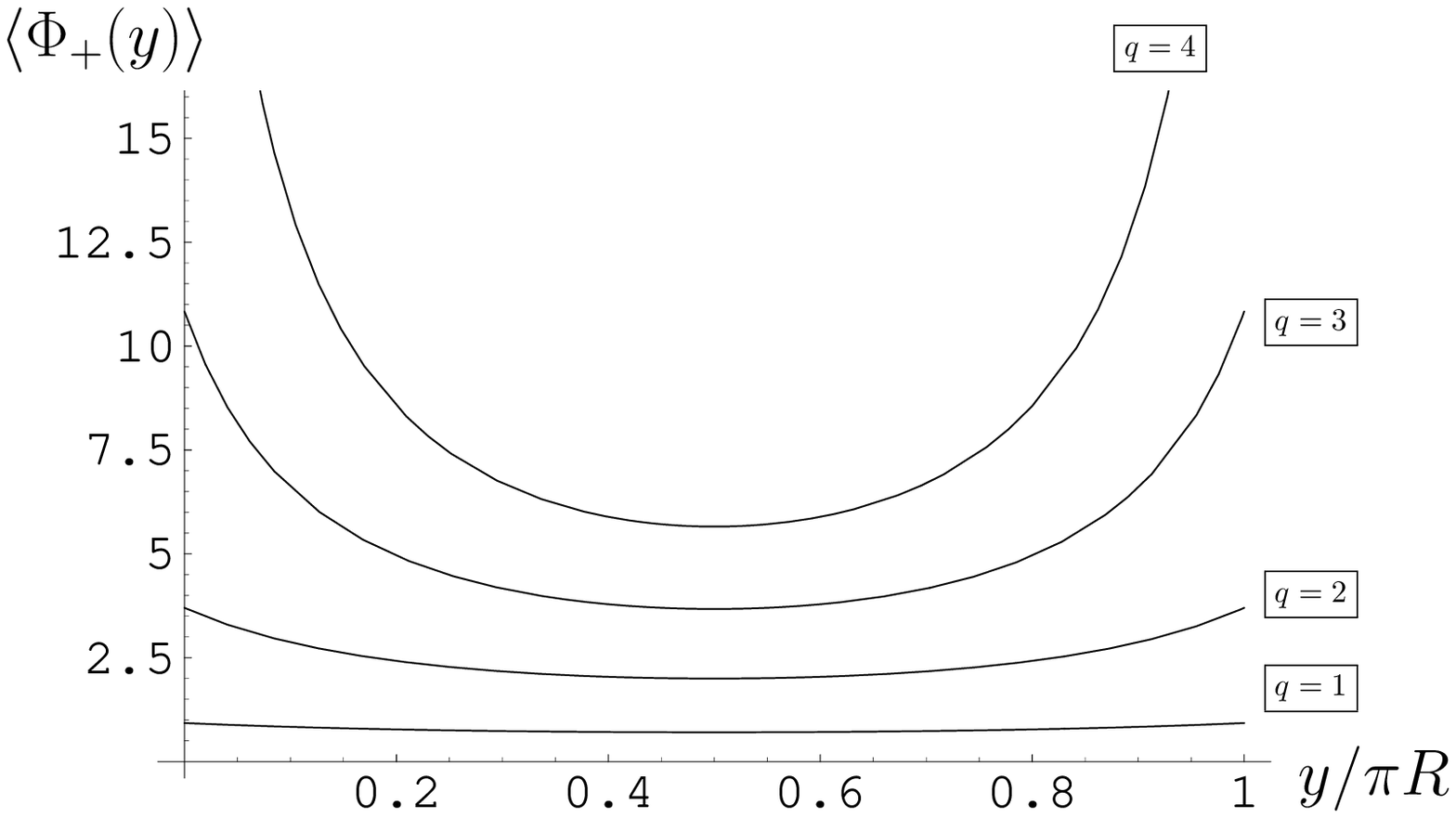,width=\linewidth}}
   \centerline{(a) $\langle \Phi_+ (y) \rangle$ $(r^\pi_0=1)$}
\end{minipage}
\hfill
\begin{minipage}{0.48\linewidth}
   \centerline{\epsfig{figure=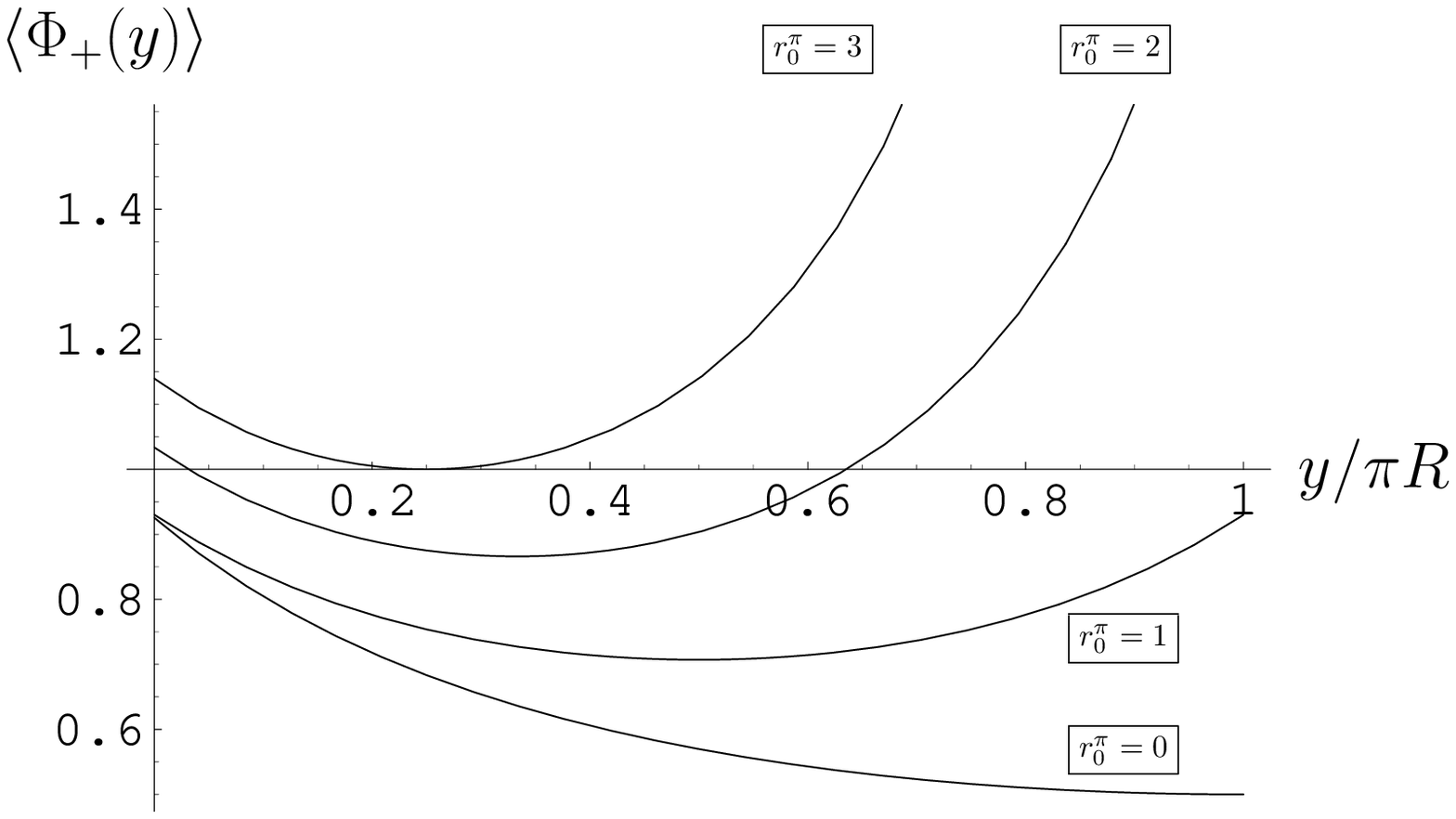,width=\linewidth}}
   \centerline{(b) $\langle \Phi_+ (y) \rangle$ $(q=1)$}
\end{minipage}
\end{center}
\begin{center}
\begin{minipage}{0.48\linewidth}
   \centerline{\epsfig{figure=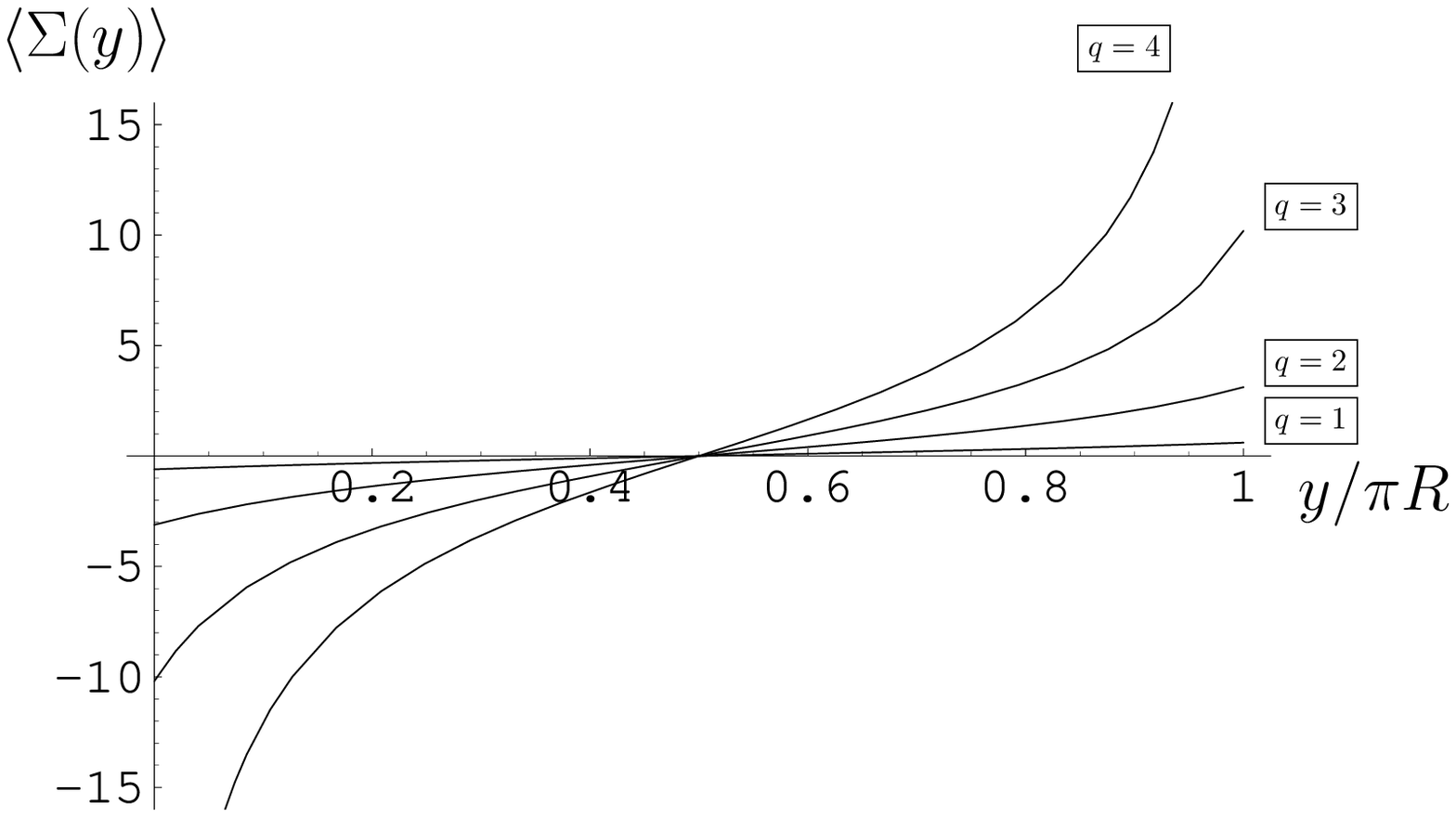,width=\linewidth}}
   \centerline{(c) $\langle \Sigma (y) \rangle$ $(r^\pi_0=1)$}
\end{minipage}
\hfill
\begin{minipage}{0.48\linewidth}
   \centerline{\epsfig{figure=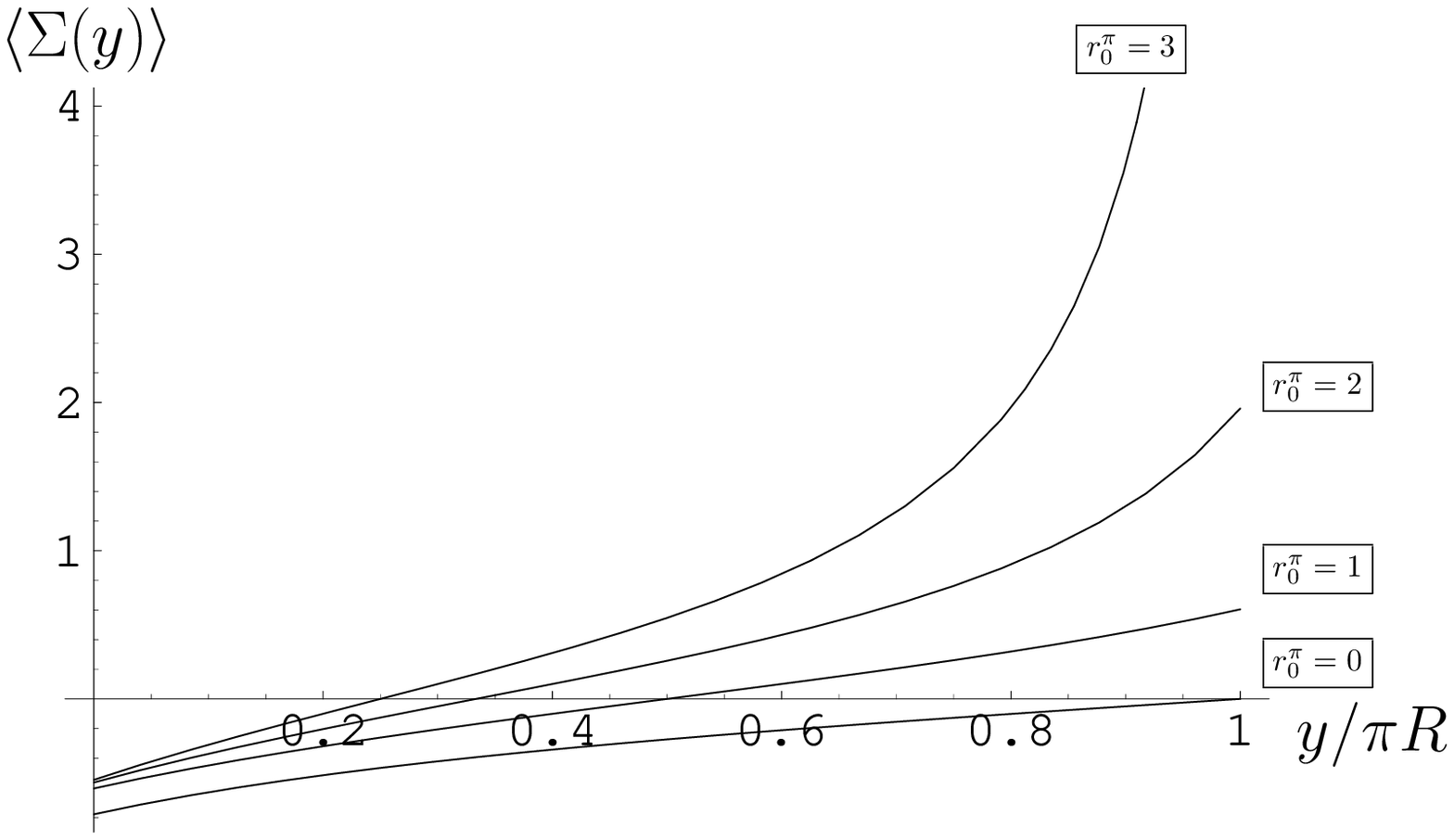,width=\linewidth}}
   \centerline{(d) $\langle \Sigma (y) \rangle$ $(q=1)$}
\end{minipage}
\end{center}
\caption{$y$ dependence of the VEVs $\langle \Phi_+(y) \rangle$ 
and $\langle \Sigma (y) \rangle$. 
These were obtained by setting $g\pi R \xi_0/2 \equiv -1$, $a=a_+$ 
($m=0$) and varying $q$ or $r^\pi_0 = \xi_\pi/\xi_0$.}
\label{fig:susybv}
\end{figure}
The boundary conditions at $y=0$ and $\pi R$ 
imply further restrictions. Explicitly, 
the boundary condition 
$\partial_y \Phi_+(0) = \Sigma(0) =0$ requires 
$b = c_\pi $,and the same condition of $y= \pi R$ 
requires 
\begin{equation}
a\pi R + c_0 + c_\pi = m \pi,
\label{sol-3}
\end{equation}
where $m$ is an integer.
We combine the latter equation in (\ref{sol-2}) and 
Eq.~(\ref{sol-3}) to obtain a further equation that must be 
satisfied by the constant $a$, and its solution is obtained as 
\begin{equation}
a_{\pm} = {m \pi \pm \sqrt{m^2\pi^2 - 2\pi R gq 
(\xi_0 + \xi_\pi)} \over 2 \pi R}.
\end{equation}
Recall that only the bulk field with charge 
$q <0$ can develop the above VEV. Thus, the value inside 
the square root must always be positive. 

Moreover we have to require the VEVs 
$\langle \Phi_+ \rangle $ and $\langle \Sigma \rangle$ to 
be non-singular in the region $0 < y < \pi R$.
This implies that the singular point 
$y=(n\pi +\pi/2-c_0)/a$,  where $n$ is any integer, 
can not be in the region 
$0 < y < \pi R$.
The VEVs of Eqs.~(\ref{vevphi-3}) and (\ref{vevSigma-3}) 
are shown in Fig.~\ref{fig:susybv}.

Now let us discuss the zero-mode profile of the bulk field 
with the charge $q'$, whose sign is opposite to that of 
the charge $q$, i.e. $qq' < 0$. 
Such a profile is obtained by substituting 
this VEV of $\Sigma$ into Eq.~(\ref{profile-1}). 
This gives 
\begin{eqnarray}
\Phi_+'^{(0)} (y) = 
\Phi_+'^{(0)} (0) \cos^{|q'/q|} [ay + b + c_0 {\rm sgn}(y) 
+ c_\pi {\rm sgn }(y-\pi R)].
\label{zerophi}
\end{eqnarray}

Next, we would like to study profiles and mass eigenvalues 
of higher modes.  Some analyses are given in the 
last part of the appendix. Here, we consider the simple case that 
argument of inside the tangent function in (\ref{vevSigma-3}) 
is sufficiently small that we can approximate 
Eq.~(\ref{vevSigma-3}) as 
\begin{equation}
\langle \Sigma (y) \rangle \approx - 
{\xi_0 + \xi_\pi \over 2 \pi R}y  + 
{\xi_0 \over 2}{\rm sgn}(y) 
+ {\xi_\pi \over 2}({\rm sgn }(y-\pi R)-1),
\label{approx-1}
\end{equation}
where we have taken $m=0$.
This VEV is exactly the same as that given in Eq.~(\ref{VEVSigma-1}).
Note that in the present case, SUSY is unbroken.
Actually, the VEV of $\Phi_+$ (\ref{vevphi-3}) is 
approximated as 
\begin{equation}
\langle |\Phi_+ (y)|^2 \rangle \approx  -
{\xi_0 + \xi_\pi \over 2 \pi R} 
\end{equation}
at the same level of approximation as Eq.~(\ref{approx-1}).
This cancels the VEV $\langle \Sigma \rangle$.
Thus, the mass eigenvalues and profiles of higher modes 
are obtained in the same way as in \S \ref{sec:2.2.1}, 
but in this case there is no SUSY breaking scalar mass terms.
Thus, the bosonic and fermionic modes have the same 
spectrum, e.g. Eq.~(\ref{fm-eigen1}).

\section{SUSY breaking from the boundary}
\label{sec:3}

To this point, we have studied VEVs and zero-mode profiles 
systematically in the cases with generic values 
of $\xi_0$ and $\xi_\pi$. Our main purpose has thus been fulfilled. 
In this section, we consider implications 
of our results for phenomenological applications. 

In any case, bulk fields have non-trivial profiles, depending 
on their $U(1)$ charges. This is useful, e.g., to explain 
hierarchical couplings, as in Refs.~\cite{KT} and \cite{Pomarol}. 
For such a purpose, in Ref.~\cite{KT} zero-mode profiles with 
the VEV (\ref{vevSigma-3}) in \S \ref{sec:2.2.3} are used, and 
in Ref.~\cite{Pomarol} the zero-mode profiles (\ref{profile-2}) 
are used. The profile (\ref{profile-3}) is also useful. 

Another result, which is important, is the $D$-term contribution 
to the SUSY-breaking scalar masses (\ref{Dterm-1}) and (\ref{Dterm-2}), 
which are obtained in \S 2.2.1. The overall magnitudes of the $D$-term 
contributions are the same, up to $U(1)$ charges, everywhere in the bulk 
and branes, because the VEV of the $D$ term is constant along the $y$ 
direction. In the case of \S 2.2.1, SUSY is broken by 
the FI terms, and its breaking scale is 
$O((g(\xi_0 + \xi_\pi)/R)^{1/2})$.
In this section, we discuss briefly $D$-term 
contributions in two other cases, that in which the $D$-flatness and 
$F$-flatness conditions are inconsistent, which leads to 
SUSY breaking, and that in which SUSY is broken dominantly by 
another mechanism independent of the FI terms.
In \S \ref{sec:3.1} we consider a toy model for SUSY breaking, 
which is a 5D simple extension of the 4D models 
studied in Ref.~\cite{SUSY-breaking:mech}.
In \S \ref{sec:3.2}, we extend the analysis of the $D$-term 
contributions in the 4D models studied in Ref.~\cite{anomalous} 
to our 5D case.

\subsection{SUSY breaking only by the brane field}
\label{sec:3.1}

Our starting point here is almost the same as that in 
\S \ref{sec:2.2.2}. We consider the case that 
all the bulk fields have positive $U(1)$ charges and 
all of the brane fields also have positive charges, 
but only one brane field $\chi_0$ at $y=0$ has 
a negative $U(1)$ charge. 
Here, we employ a normalization such that this field has the 
$U(1)$ charge $q_{\chi_0} = -1$. 
If this brane field has no superpotential 
on the brane, there exists the SUSY-flat direction 
studied in \S \ref{sec:2.2.2}, i.e., 
\begin{equation}
\partial_y \langle \Sigma \rangle = 
(\xi_0 - g\langle |\chi_0|^2\rangle )\delta (y) + 
\xi_\pi \delta (y-\pi R), \qquad 
\langle |\chi_0|^2\rangle   = {\xi_0 + \xi_\pi \over g}.
\end{equation}
Along this direction, $U(1)$ is broken, although SUSY is unbroken.

Here we assume the mass term of $\chi_0$ 
in the superpotential to be 
\begin{equation}
W = m\chi_0 \chi'_0 \delta (y),
\label{mass-term}
\end{equation}
where $\chi'_0$ is the brane field on $y=0$ 
and has $U(1)$ charge $q_{\chi'_0} =1$.
This type of mass term can be generated 
by dynamics on the brane, as discussed in the case of 4D models 
in Refs.~\cite{SUSY-breaking:mech} and \cite{SUSY-breaking2}.
In this case, the dynamically generated mass can 
also have a $U(1)$ charge, and other types of 
$U(1)$ charge assignments for $\chi'_0$ are 
possible. Alternatively, we could write the mass term in terms of 
$\chi_0$ alone. For the sake of simplicity of our toy model, 
we use the above mass term and charge assignment.

With this mass term, SUSY is broken.
Suppose that the field $\chi_0$ develops a VEV 
$\langle \chi_0 \rangle = v $. 
Then the analysis in \S \ref{sec:2.2.1} implies that 
the scalar potential $V$ is minimized by 
the following VEV of $\Sigma$: 
\begin{eqnarray}
\partial_y \langle \Sigma \rangle &=& \xi(y) 
-gv^2 \delta(y) +C,\\
\langle \Sigma \rangle &=& {1\over 2}(\xi_0 - gv^2){\rm sgn} (y) 
+ Cy + {1\over 2}\xi_\pi({\rm sgn} \left(y-\pi R)+1\right),
\end{eqnarray}
where 
$C = - ({\xi_0 -gv^2 + \xi_\pi)/(2 \pi R)}$.
The 4D scalar potential with 
the above brane mass term (\ref{mass-term}) 
is written 
\begin{equation}
V_4 = m^2v^2 + {1 \over 2} 
{(\xi_0 +\xi_\pi - gv^2)^2 \over 2\pi R}.
\end{equation}
Minimizing this potential, we obtain $v$ as 
\begin{equation}
v^2 = {\xi_0 + \xi_\pi \over g} - 
{ 2 \pi R \over g^2}m^2 .
\end{equation}
With this VEV, the following SUSY-breaking parameters are obtained: 
\begin{eqnarray}
-\partial_y \langle \Sigma \rangle + \xi (y) 
-g \langle |\chi_0|^2 \rangle = 
{m^2 \over g}, \label{vevSigma-5}\\
\left|F_{\chi'_0}\right| = m v  = 
m\sqrt{{\xi_0 + \xi_\pi \over g} - 
{ 2 \pi R \over g^2}m^2 }.
\end{eqnarray}
Note that this $F$-term is induced only on the 
$y=0$ brane.

With this VEV of $\Sigma$, the fermionic 
zero mode as well as the corresponding bosonic mode 
has the profile 
\begin{equation}
\Phi_+^{(0)}(y) = A \exp \left[-{1 \over 2}qm^2 
\left(y- {g\xi_0 \over 2m^2} \right)^2 \right]. 
\label{profile-5}
\end{equation}
That is the Gaussian profile. 
Using a method similar to that of \S \ref{sec:2.2.1},
 profiles of higher modes as well as 
their mass eigenvalues are obtained.

In this vacuum, SUSY is broken and 
SUSY breaking terms are induced by 
non-vanishing $D$ and $F$ terms. 
All of the bulk gaugino masses are induced by 
\begin{equation}
\int  d\theta^2 {\chi_0 \chi'_0 \over \Lambda^3} 
W^\alpha W_\alpha \delta(y),
\end{equation}
where $\Lambda$ is the cut-off scale; 
that is, a zero-mode gaugino mass $M_{\lambda_0}$ is induced as 
\begin{equation}
M_{\lambda_0} \sim {F_{\chi'_0} \chi_0 \over 2 \pi R \Lambda^3} =
{m v^2 \over 2 \pi R \Lambda^3} .
\end{equation}
The bulk field with charge $q$ 
has a SUSY-breaking scalar mass due to non-vanishing 
$D$ term given by 
\begin{equation}
m^2_q  = qm^2 .
\end{equation}
The brane field has the same $D$-term contribution 
to the SUSY breaking scalar mass, $m^2_I = q_I m^2 $ .
Here, the VEV of the $\Sigma$ field plays a role to 
mediate SUSY breaking to bulk fields and brane fields 
on the other brane as $D$-term contributions to 
SUSY breaking scalar masses.

The brane field on the $y=0$ brane 
has the contact term 
\begin{equation}
\int d^4\theta{|\chi_0|^2 \over \Lambda^2} |\phi_0|^2, 
\end{equation}
which induces the following contribution to scalar masses 
\begin{equation}
m^2_{\phi_0} \sim {m^2 v^2 \over \Lambda^2}  .
\end{equation}
Similarly, the bulk scalar fields with vanishing 
$U(1)$ charge have a scalar mass $m^2_{\Phi} \sim
m^2 v^2 / (R \Lambda^3)$ from the $y=0$ brane.

In the case $R = O(1/\Lambda)$, we have the relation 
\begin{equation}
m_q : m_{\phi_0} : |M_\lambda| = m : m\varepsilon : 
m \varepsilon^2,
\end{equation}
where $\varepsilon = v/\Lambda$.
If $R$ is larger than $O(1/\Lambda)$, we would have much smaller gaugino 
masses than scalar masses.
Also, note that there are many Kaluza-Klein modes below the cutoff
scale in the case $R\Lambda > O(10)$.
Thus, for such case radiative corrections from Kaluza-Klein modes to 
gauge couplings \cite{Dienes:1998vg},
Yukawa couplings \cite{Bando:2000it} and SUSY breaking 
terms \cite{Kobayashi:1998ye,Kubo:2002pv} may be important.
As stated above, the mass spectrum of higher modes 
is obtained similarly to that of \S 2.2.1. 
Here we do not study the details of this procedure 
because this is not our purpose.

We have a $D$-term contribution to scalar masses, as in 4D theory.
An important point is that the $D$ term is constant along 
the $y$ direction in our toy model.
Therefore, bulk fields and brane fields on both branes have 
the same effects. The field $\Sigma$ plays a role with regard to 
this point. Its VEV, in particular $\partial_y \langle \Sigma \rangle$, 
levels (flattens) the localized $D$ term in the bulk. 
This is the reason that the $D$-term contribution appears 
everywhere in the bulk. 
Therefore, $D$-term contributions, in general, generate 
charge-dependent sfermion masses everywhere 
in bulk and on both branes. 

A bulk field could play the same role as that of the brane 
field $\chi'_0$, and we would obtain the same result. 
In this case we could also assume 
that the mass term superpotential exists only on the $y=0$ brane. 
We could consider more complicated models, e.g., a model 
with two $U(1)$s and their FI terms. 
Such cases will be studied elsewhere, with the purpose of  
interesting aspects, e.g., realistic Yukawa couplings 
and sparticle spectra.

\subsection{$D$-term contribution in the 5D model}
\label{sec:3.2}

In the above, we have assumed that the superpotential (\ref{mass-term})
breaks SUSY with a combination effect of the FI terms.
Here, we give a comment on other cases without 
such a specific superpotential.
With the same system, except the superpotential (\ref{mass-term}), 
we assume that SUSY is softly broken dominantly 
through some mechanism and a SUSY-breaking scalar mass $m_{\chi_0}$ 
of $\chi_0$ is induced.
Through this SUSY breaking, gaugino masses and soft scalar masses 
of other fields can be induced. In this case, the previous analysis holds, 
if $m$ is replaced by $m_{\chi_0}$. 
The $\Sigma$ field has a non-trivial VEV, as given in Eq.~(\ref{vevSigma-5}), 
and the fermionic zero mode of bulk field has a non-trivial profile, 
as given in Eq.~(\ref{profile-5}).
Furthermore, bulk fields with charge $q$ and 
brane fields with charge $q_I$ on both branes 
have $D$-term contributions to scalar masses given by 
\begin{equation}
m^2_q = q m^2_{\chi_0},\qquad m^2_I = q_I m^2_{\chi_0},
\end{equation}
respectively. Again, the VEV of the $\Sigma$ field plays a role to 
induce $D$-term contributions to 
SUSY breaking scalar masses
for bulk fields and brane fields on the other brane.

\section{Conclusion and discussion}
\label{sec:4}

We have systematically studied 
VEVs of $\Sigma$ and brane/bulk fields 
in a 5D model with generic FI terms. 
We have considered three cases: i) the case 
in which non-vanishing FI terms are not cancelled by the 
VEVs of brane/bulk scalar fields, ii) the case in which 
FI terms are cancelled by the VEV of the brane field, and 
iii) the case in which FI terms are cancelled by the 
VEV of bulk fields.  
The non-trivial VEV of $\Sigma$ generates bulk mass terms for 
$U(1)$ charged fields, and their zero modes have 
non-trivial profiles. 
In the first case, SUSY is broken, and 
fermionic zero modes as well as the corresponding bosonic modes 
have Gaussian profiles. 
In the second case, bulk fields have exponential profiles.
In the third case, bulk fields have profiles of the form 
(\ref{zerophi}).
We also studied profiles and mass eigenvalues of higher modes.
These are our main results.
Non-trivial profiles would be useful to explain 
hierarchical couplings.

We also studied a toy model for SUSY breaking, and we obtained 
sizable $D$-term contributions to scalar masses. 
The overall magnitude of the $D$-term contributions 
are same everywhere in bulk and also on both branes.
We have to take into account these $D$-term contributions and 
other SUSY-breaking terms in a realistic model.

Our analysis is purely classical. 
Actually, one-loop calculations were carried out to show that 
the FI term is generated at the one-loop level. 
In the generic case, $U(1)$ is anomalous.
For quantum consistency, we would need 
anomaly cancellation by the Green-Schwarz mechanism and/or 
the Chern-Simons term.
For the former, brane/bulk singlet (moduli) fields 
might be transformed as Eq.~(\ref{GS-S}), 
so that the Green-Schwarz mechanism is effective. 
However, this is not clear.
For the latter, the SUSY Chern-Simons term includes 
the term $\Sigma F^{MN} F_{MN}$.\cite{AGW} 
(See also Refs.~\cite{Arkani-Hamed:2001is} and \cite{Pilo:2002hu}.)
Thus, a non-trivial VEV of $\Sigma$ might change 
the gauge coupling in the bulk and/or brane, although in most cases 
$U(1)$ is broken.

Nontrivial profiles of bulk fields due to $\langle \Sigma \rangle $ 
might generate FI terms at the one-loop level that are 
different from that generated in the case with 
$\langle \Sigma \rangle =0$.
For example, in the case with $\langle \Sigma \rangle =0$, 
the bulk field contributes both to $\xi_0$ and $\xi_\pi$ 
with the same weight.\cite{SSSZ,Nilles,Pomarol} 
However, suppose the extreme case in which 
a non-trivial VEV $\langle \Sigma \rangle $ 
causes the profile of bulk field to be localized completely on 
the $y=0$ brane. Then, such a bulk field might contribute to 
the FI coefficient only on the $y=0$ brane. 
We will study this type of consistency elsewhere.

Also, one-loop calculations show that FI terms can 
include second derivatives of delta functions, $\delta''(y)$ and 
$\delta''(y-\pi R)$.\cite{SSSZ,Nilles,Pomarol} 
Such terms drastically change the behavior of profiles 
on the branes.\cite{Nilles} 
We can extend our analysis to the case including 
second derivatives of delta functions.

\section*{Acknowledgements}
The authors would like to thank Stefan~Groot~Nibbelink, 
Hiroyuki~Hata and Hiroaki~Nakano for fruitful discussions and 
correspondences. 
We also would like to thank the organizers of Summer 
Institute 2002 at Yamanashi, Japan. 
T.~K. is supported in part by, Grant-in-Aid for 
Scientific Research from the Ministry of Education, Science, 
Sports and Culture of Japan (\#14540256).

\appendix
\section*{Higher modes}

Here we study in detailed profiles and mass eigenvalues 
of higher modes with the non-trivial VEV of $\Sigma$ 
obtained in \S 2.2.1 and 2.2.3.

\subsection*{The case with $\langle \Sigma\rangle  \neq 0$ 
($^\forall q, q_I>0$)}
First we study the case of \S 2.2.1.
{}From the potential (\ref{potential}) we have a $\Phi_\pm$ square 
term after $\Sigma$ develops the VEV $\langle \Sigma \rangle$. 
It is written 
\begin{eqnarray}
-{\cal L}_{\rm m}^{\Phi_\pm} &=& 
   \pm \left( \xi(y)-\partial_y \langle \Sigma \rangle \right) 
   gq |\Phi_\pm|^2 
 + |\partial_y \Phi_\pm \mp gq \langle \Sigma \rangle \Phi_\pm|^2
\nonumber \\ &=& 
-\Phi_\pm^\dagger \left( \partial_y^2 \mp gq \xi (y) 
 - (gq)^2 \langle \Sigma \rangle^2 \right) \Phi_\pm 
\equiv 
-\Phi_\pm^\dagger \Delta_\pm \Phi_\pm, 
\label{KKop}
\end{eqnarray}
where $\Delta_\pm \equiv \partial_y^2 \mp gq \xi(y) 
- (gq)^2 \langle \Sigma \rangle^2$. The eigenvalues of the operator 
$\Delta_\pm$ correspond to the KK spectrum of $\Phi_\pm$. 

We solve the eigenvalue equation 
\begin{eqnarray}
\Delta_\pm \Phi_\pm + \lambda \Phi_\pm =0, 
\label{PhiEE}
\end{eqnarray}
where $\lambda$ is the eigenvalue of the operator $\Delta_\pm$. 
Following Ref.~\cite{Nilles}, we redefine $\Phi_\pm$ as 
\begin{eqnarray}
\Phi_\pm = 
\widetilde\Phi_\pm e^{\pm gq \int_0^y dy' \langle \Sigma(y') \rangle},
\label{redeftp}
\end{eqnarray}
and substitute this into Eq.~(\ref{PhiEE}). This yields the eigenvalue 
equation for $\widetilde\Phi_\pm$
\begin{eqnarray}
\left( \partial_y^2 \pm 2gq \langle \Sigma \rangle \partial_y 
\pm gq \left\{ \partial_y \langle \Sigma \rangle - \xi(y) \right\} 
+ \lambda \right) \widetilde\Phi_\pm =0, 
\label{tPhiEE}
\end{eqnarray}
with the boundary conditions 
\begin{eqnarray}
\partial_y \widetilde\Phi_+(y)\Big|_{y=0,\pi R}=
\widetilde\Phi_-(y)\Big|_{y=0,\pi R}=0. 
\label{tPhiBC}
\end{eqnarray}

The VEVs $\partial_y \langle \Sigma \rangle$ and 
$\langle \Sigma \rangle$ are given by 
Eqs.~(\ref{vevphi}) and (\ref{VEVSigma-1}), respectively. 
In the bulk ($0<y<\pi R$), the eigenvalue equation (\ref{tPhiEE}) becomes 
\begin{eqnarray}
\partial_y^2 \widetilde\Phi_\pm 
\pm 2gq \left( Cy+{\xi_0 \over 2}\right) \partial_y \widetilde\Phi_\pm 
 +  \left( \lambda \pm gqC \right) \widetilde\Phi_\pm =0.
\label{bulkEE}
\end{eqnarray}

In the case $\xi_0 + \xi_\pi = C =0$, 
Eq.~(\ref{bulkEE})  is solved in Ref.~\cite{Nilles}.
The (mass)$^2$ eigenvalues $\lambda_k$ and wave functions 
of higher modes for $Z_2$ even bulk fields are obtained as 
\begin{eqnarray}
\lambda_k &=& {k^2 \over R^2} + {(gq\xi_0)^2 \over 4}, \\
\tilde \Phi^{(k)}_+ &=& \alpha_- e^{\alpha_+ y} -
\alpha_+ e^{\alpha_- y},
\end{eqnarray}
up to normalization, where 
\begin{equation}
\alpha_\pm = - {1 \over 2}gq\xi_0 \pm {k \over R }i.
\end{equation}
$Z_2$ odd bulk fields have the same (mass)$^2$ eigenvalues 
$\lambda_k$, and the corresponding wave functions are obtained as 
\begin{equation}
\tilde \Phi^{(k)}_- = e^{-\alpha_+ y} - e^{-\alpha_- y},
\end{equation}
up to normalization.

\begin{figure}[t]
\begin{center}
\begin{minipage}{0.48\linewidth}
   \centerline{\epsfig{figure=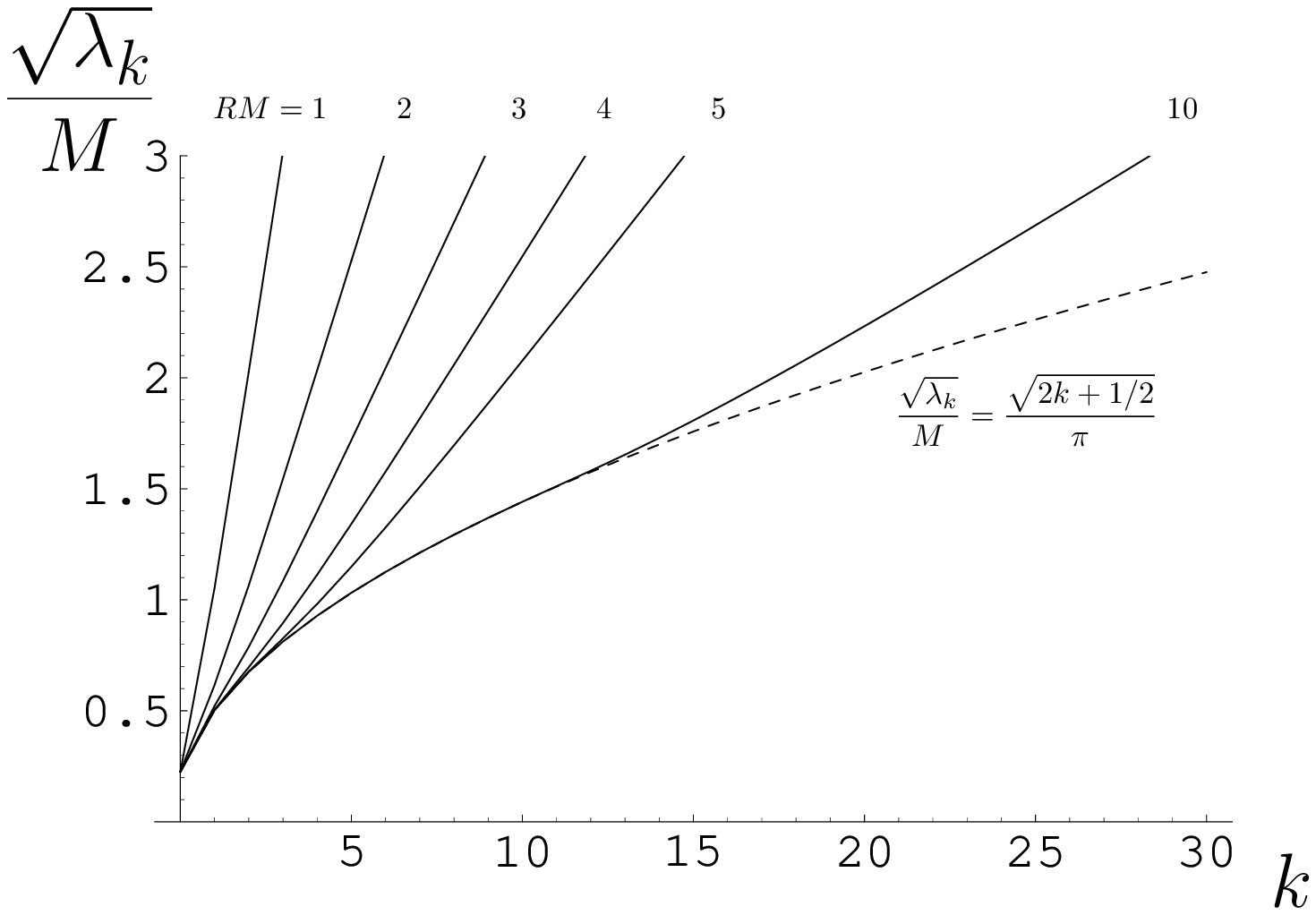,width=\linewidth}}
   \centerline{(a) $\Phi_+$, $z_0=RM$, $z_\pi=0$}
\end{minipage}
\hfill
\begin{minipage}{0.48\linewidth}
   \centerline{\epsfig{figure=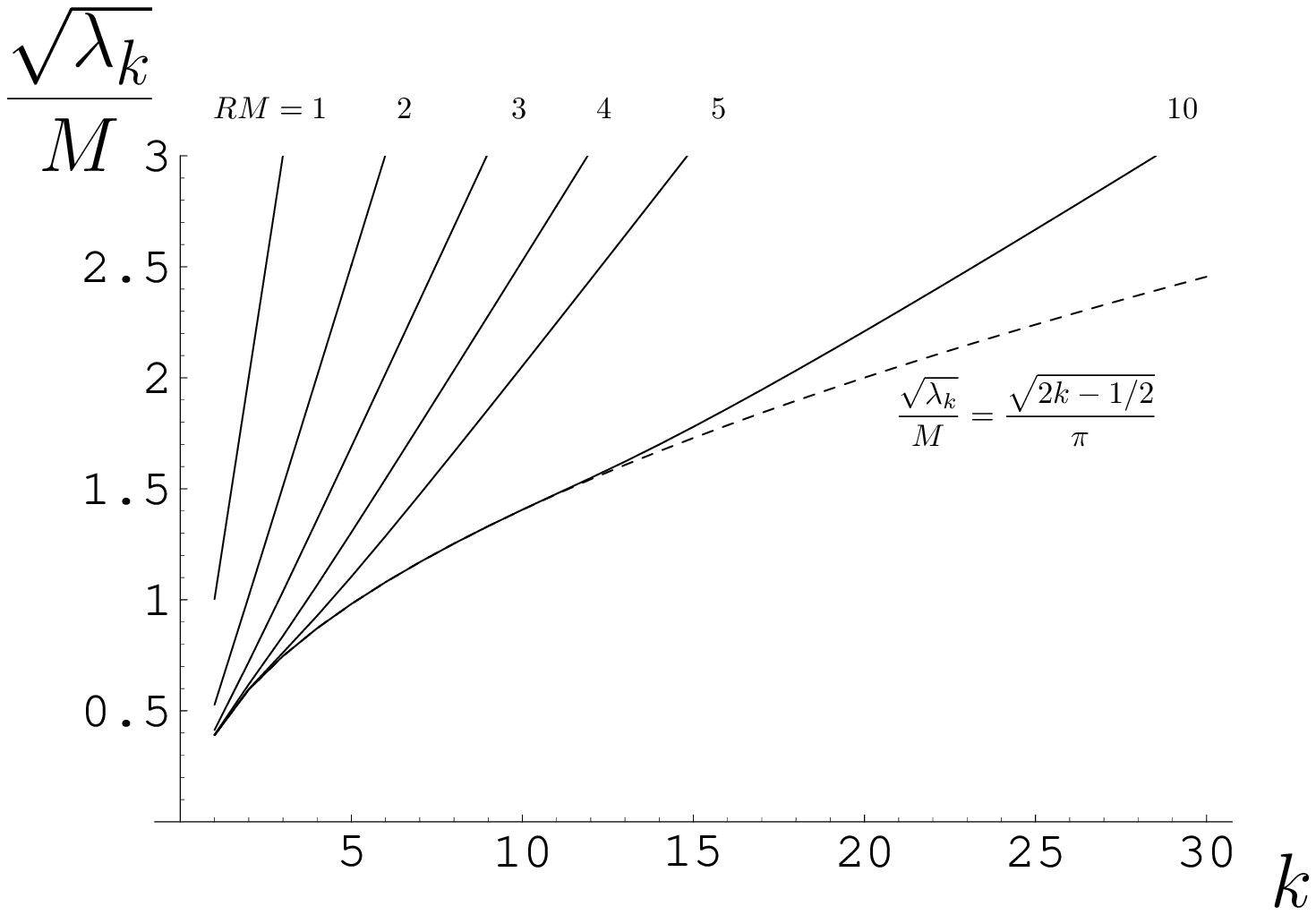,width=\linewidth}}
   \centerline{(b) $\Phi_-$, $z_0=RM$, $z_\pi=0$}
\end{minipage}
\end{center}
\begin{center}
\begin{minipage}{0.48\linewidth}
   \centerline{\epsfig{figure=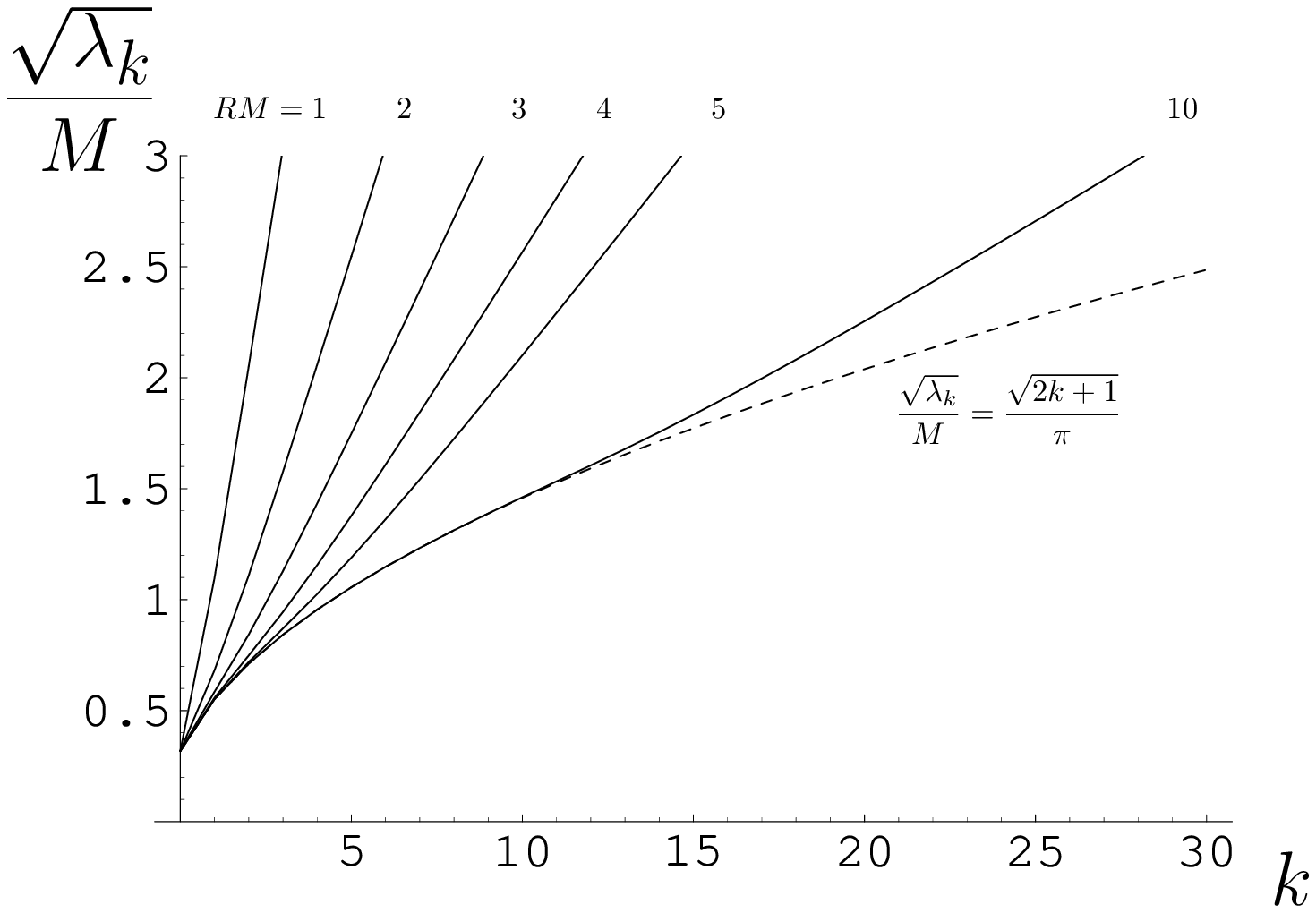,width=\linewidth}}
   \centerline{(c) $\Phi_+$, $z_0=z_\pi=RM/\sqrt{2}$}
\end{minipage}
\hfill
\begin{minipage}{0.48\linewidth}
   \centerline{\epsfig{figure=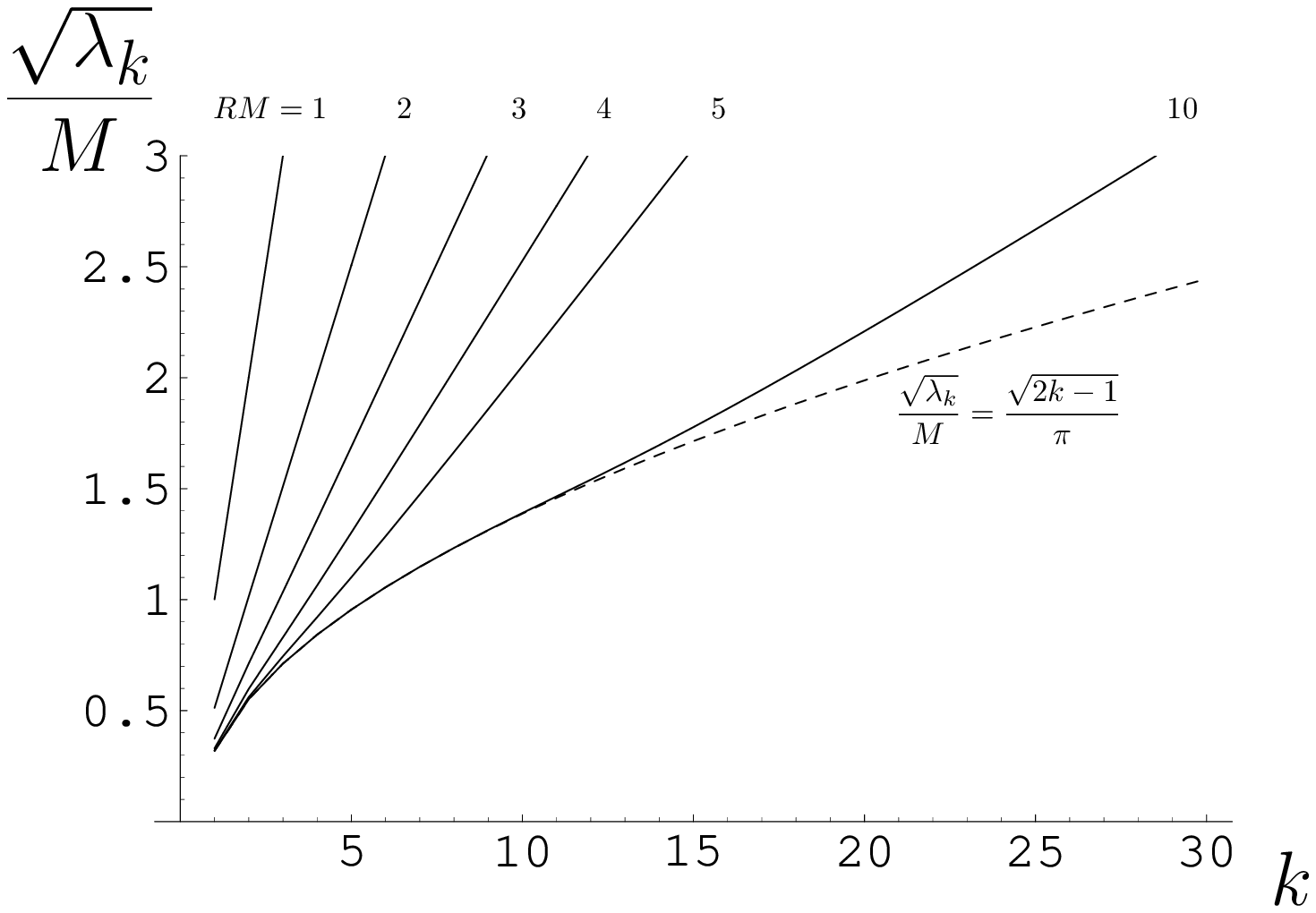,width=\linewidth}}
   \centerline{(d) $\Phi_-$, $z_0=z_\pi=RM/\sqrt{2}$}
\end{minipage}
\end{center}
\caption{KK spectrum $\sqrt{\lambda_k}/M$ up to the 30th excited mode 
for the case $z_0=RM$ and $z_\pi=0$ 
($l_{\xi_0}=1/(4\pi^2g_4^2q)$ and $l_{\xi_\pi}=0$) in (a) and (b),  
and for the case $z_0=z_\pi=RM/\sqrt{2}$ 
($l_{\xi_0}=l_{\xi_\pi}=1/(4\pi^2g_4^2q)$) in (c) and (d). 
The discrete eigenvalues are joined by solid curves. 
{}From (a) and (b), we see that 
$\lambda_k$ 
$= \left( \frac{M}{\pi} \right)^2 
\left\{ \left( \frac{\pi k}{RM} \right)^2 \pm \frac{1}{2} \right\}$ 
($k \gg RM$), and 
$\lambda_k$ 
$= \left( \frac{M}{\pi} \right)^2 \left(2k\pm \frac{1}{2} \right)$ 
($k \ll RM$). 
{}From (c) and (d), we see that 
$\lambda_k$ 
$= \left( \frac{M}{\pi} \right)^2 
\left\{ \left( \frac{\pi k}{RM} \right)^2 \pm 1 \right\}$ 
($k \gg RM$), and 
$\lambda_k$ 
$= \left( \frac{M}{\pi} \right)^2 (2k\pm 1)$ 
($k \ll RM$). 
The case $k \ll RM$ is rather non-trivial, because $\lambda_k$ 
is linear in $k$, and 
it does not depend on $R$, the size of extra dimension. 
This can be understood from the fact that Eq.~(\ref{tPhiEEz}) 
is a Hermite differential equation (harmonic oscillator) 
for $R \to \infty$. 
We also note that the first excited mode of $\Phi_-$ 
does not become tachyonic, because 
$\lambda_k/M^2 \ge (2k-1)/\pi^2$, from (b) and (d).}
\label{fig:sp1}
\end{figure}

In the case $\xi_0 + \xi_\pi = C \ne 0$, 
the differential equation (\ref{bulkEE}) can be simplified as 
\begin{eqnarray}
\partial_z^2 \widetilde\Phi_\pm (z) 
\mp z \partial_z \widetilde\Phi_\pm (z)
+ \nu \widetilde\Phi_\pm (z) = 0, 
\label{tPhiEEz}
\end{eqnarray}
where $z=(Cy+{\xi_0 \over 2})\sqrt{\frac{2gq}{-C}}$ and 
$\nu=\frac{\lambda}{-2gqC}\mp \frac{1}{2}$. 
The boundary conditions (\ref{tPhiBC}) become 
\begin{eqnarray}
\partial_z 
\widetilde\Phi_+(z)\Big|_{z=z_0,z_\pi}= 
\widetilde\Phi_-(z)\Big|_{z=z_0,z_\pi}=0, 
\label{tPhiBCz}
\end{eqnarray}
where $z_0={\xi_0 \over 2}\sqrt{\frac{2gq}{-C}}$ and 
$z_\pi=-\frac{\xi_\pi}{2}\sqrt{\frac{2gq}{-C}}$. 
By parameterizing 
\begin{eqnarray}
\xi_I = 2g  l_{\xi_I} M^2, 
\end{eqnarray}
where $I=(0,\pi)$, 
$l_{\xi_I}$ is a dimensionless parameter, 
and $M$ is a representative scale for the FI terms, 
we obtain the boundary parameters $z_0$ and $z_\pi$ as 
\begin{eqnarray}
z_0 = 
\frac{2\pi g_4 \sqrt{q}  l_{\xi_0}}
{\sqrt{ l_{\xi_0}+ l_{\xi_\pi}}}
RM, \quad 
z_\pi =  -r^\pi_0 z_0,
\label{defz}
\end{eqnarray}
where $r^\pi_0 =  l_{\xi_\pi} /  l_{\xi_0} 
=\xi_\pi / \xi_0$. 
The solution of Eq.~(\ref{tPhiEEz}) 
is given by 
\begin{eqnarray}
\widetilde\Phi_\pm (z) = f_1^\pm F_1^\pm (z,\nu) + f_2^\pm F_2^\pm (z,\nu), 
\end{eqnarray}
where $f_1^\pm$ and  $f_2^\pm$ are arbitrary constants and 
\begin{eqnarray}
F_1^\pm (z,\nu) &=& F(\mp \nu/2\,;\,1/2\,;\,\pm z^2/2), \\
F_2^\pm (z,\nu) &=& z F(1/2 \mp \nu/2\,;\,3/2\,;\,\pm z^2/2). 
\end{eqnarray}
Here, $F(a\,;\,b\,;\,c)$ is Kummer's hypergeometric function. 
Requiring that $f_1^\pm$ and $f_2^\pm$ cannot be zero 
simultaneously in the boundary condition 
(\ref{tPhiBCz}), the following equations should be satisfied: 
\begin{eqnarray}
\partial_z F_1^{+}(z,\nu)\Big|_{z=z_0} 
\partial_z F_2^{+}(z,\nu)\Big|_{z=z_\pi} 
&=& 
\partial_z F_1^{+}(z,\nu)\Big|_{z=z_\pi}
\partial_z F_2^{+}(z,\nu)\Big|_{z=z_0}, 
\nonumber \\
F_1^-(z,\nu)\Big|_{z=z_0}F_2^-(z,\nu)\Big|_{z=z_\pi} 
&=& F_1^-(z,\nu)\Big|_{z=z_\pi}F_2^-(z,\nu)\Big|_{z=z_0}. 
\label{KKEE}
\end{eqnarray}
These determine the eigenvalue $\nu$, which also gives $\lambda$. 
We numerically solved Eq.~(\ref{KKEE}) and evaluate the eigenvalue 
for the typical case shown in Fig.~\ref{fig:sp1}. 
{}From this figure, in the case $r^\pi_0=0$, 
we see that for $\Phi_{\pm}$, $m_k^2 \equiv \lambda_k$ behaves as 
\begin{eqnarray}
\begin{array}{llll}
\lambda_k & \sim & \left( \frac{M}{\pi} \right)^2 
\left\{ \left( \frac{\pi k}{RM} \right)^2 \pm \frac{1}{2} \right\}, 
& (k \gg RM) \\
\lambda_k & \sim & \left( \frac{M}{\pi} \right)^2 (2k\pm \frac{1}{2}), 
& (k \ll RM) 
\end{array}
\end{eqnarray}
and in the case $r^\pi_0=1$, we obtain for $\Phi_{\pm}$, 
\begin{eqnarray}
\begin{array}{llll}
\lambda_k & \sim & \left( \frac{M}{\pi} \right)^2 
\left\{ \left( \frac{\pi k}{RM} \right)^2 \pm 1 \right\}, & (k \gg RM) \\
\lambda_k & \sim & \left( \frac{M}{\pi} \right)^2 (2k\pm 1), & (k \ll RM) 
\end{array}
\end{eqnarray}
with the fixed parameter 
$\xi_0/\sqrt{\pi R} \equiv M^2/(\sqrt{2}\pi^2 g_4 q)$ in all cases. 
The case with $k \ll RM$ is rather non-trivial because $\lambda_k$ 
is linear in $k$ and it does not depend on $R$, the size of 
the extra dimension. 
This behavior can be understood as follows. Equation~(\ref{tPhiEEz}) 
is a Hermite differential equation (harmonic oscillator) 
for $R \to \infty$. 
Thus, we have a spectrum like that of a harmonic oscillator for $k\ll RM$.
On the other hand, when $\lambda$ is large compared with 
$g\xi_0/R$ and $gC$, the term with 
$\partial_y \tilde \Phi_{\pm}$ in Eq.~(\ref{bulkEE}) can be ignored.
Thus, we would have $\lambda \sim (k/R)^2$.

\begin{figure}[t]
\begin{center}
$\left\{ 
\begin{array}{l}
z_0=1 \\ z_\pi=0 
\end{array}
\right.$
\hfill
\begin{minipage}{0.43\linewidth}
   \centerline{\epsfig{figure=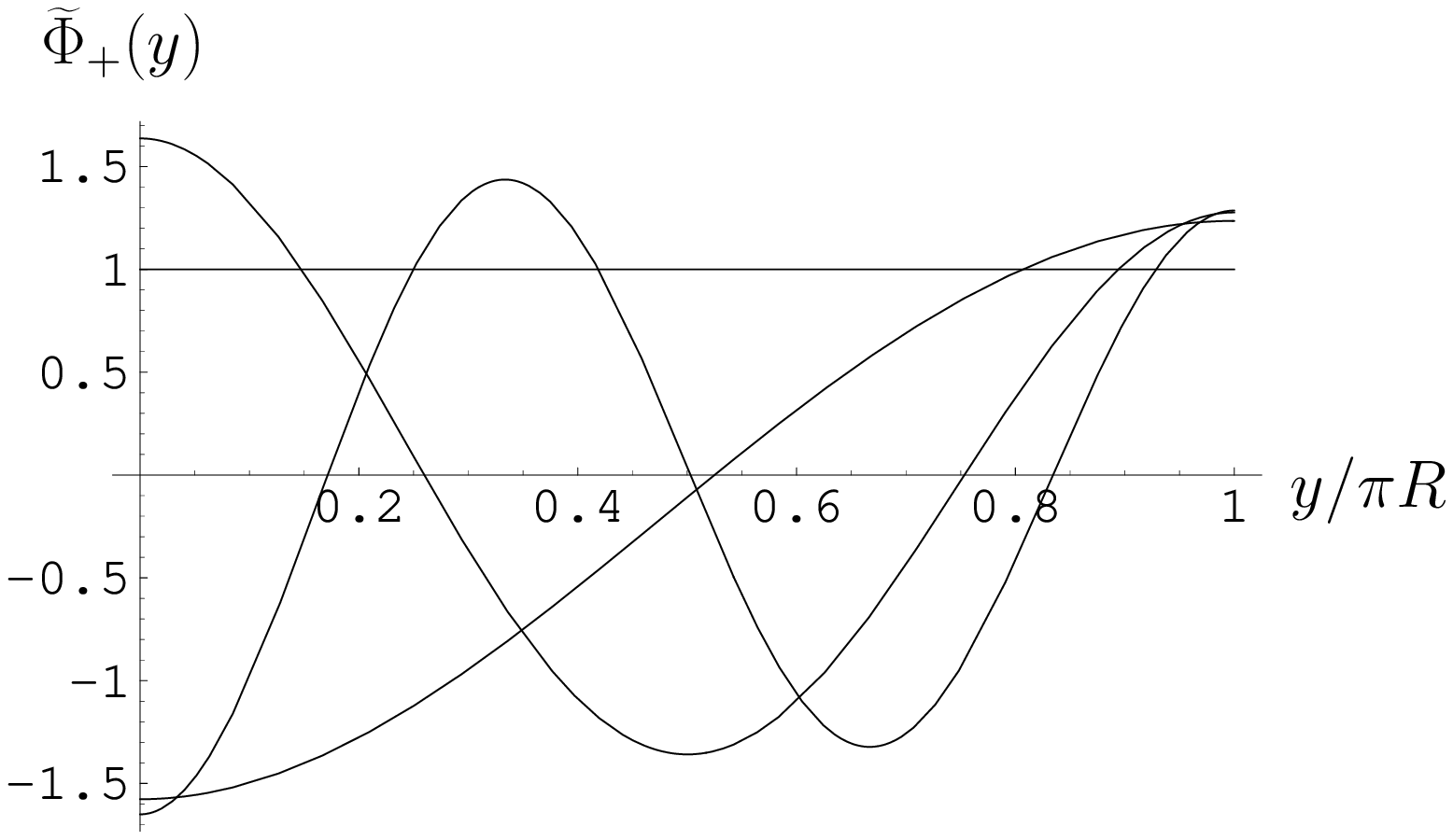,width=\linewidth}}
\end{minipage}
\hfill
\begin{minipage}{0.43\linewidth}
   \centerline{\epsfig{figure=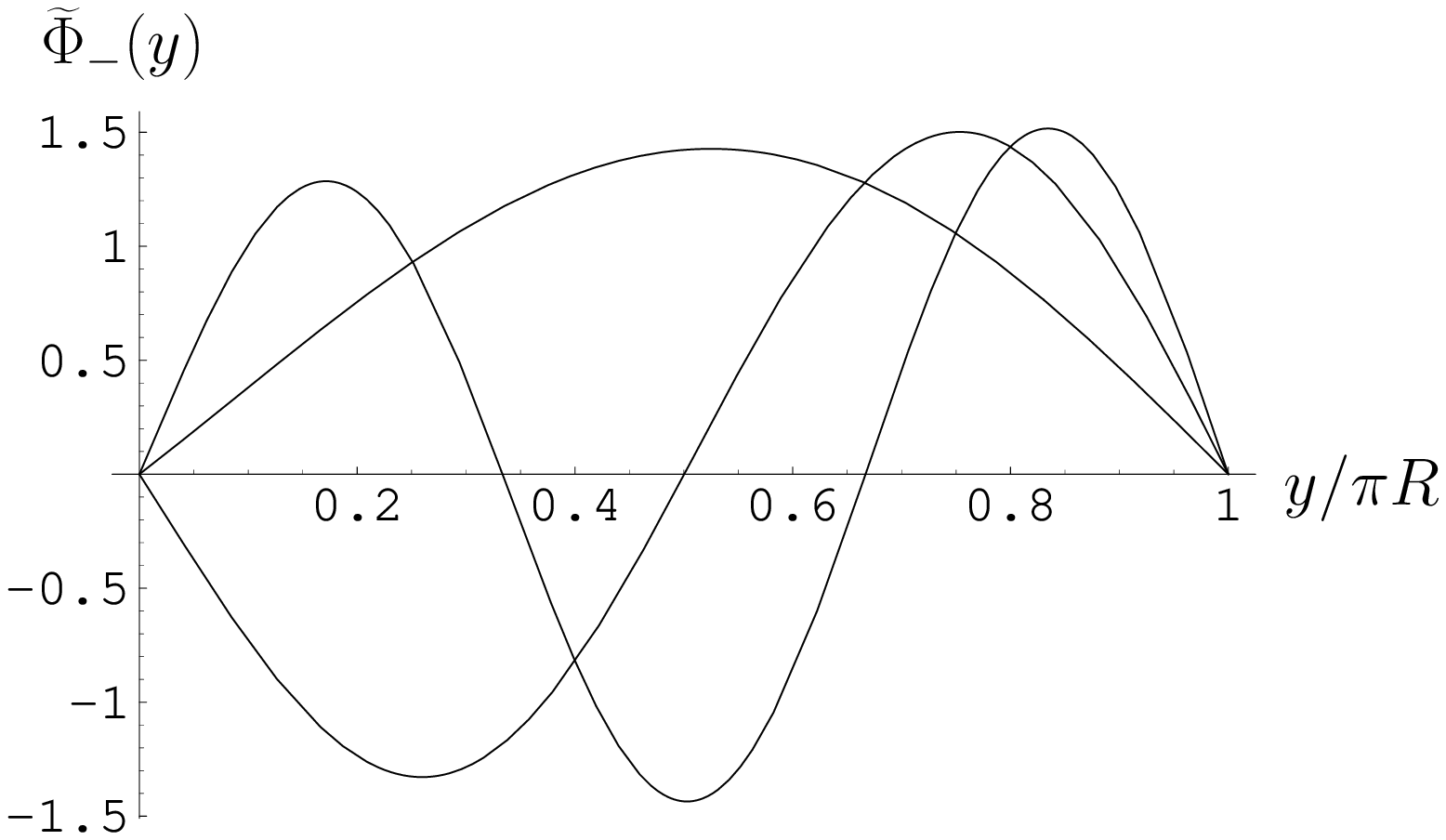,width=\linewidth}}
\end{minipage}
\end{center}
\begin{center}
$\left\{ 
\begin{array}{l}
z_0=3 \\ z_\pi=0
\end{array}
\right.$
\hfill
\begin{minipage}{0.43\linewidth}
   \centerline{\epsfig{figure=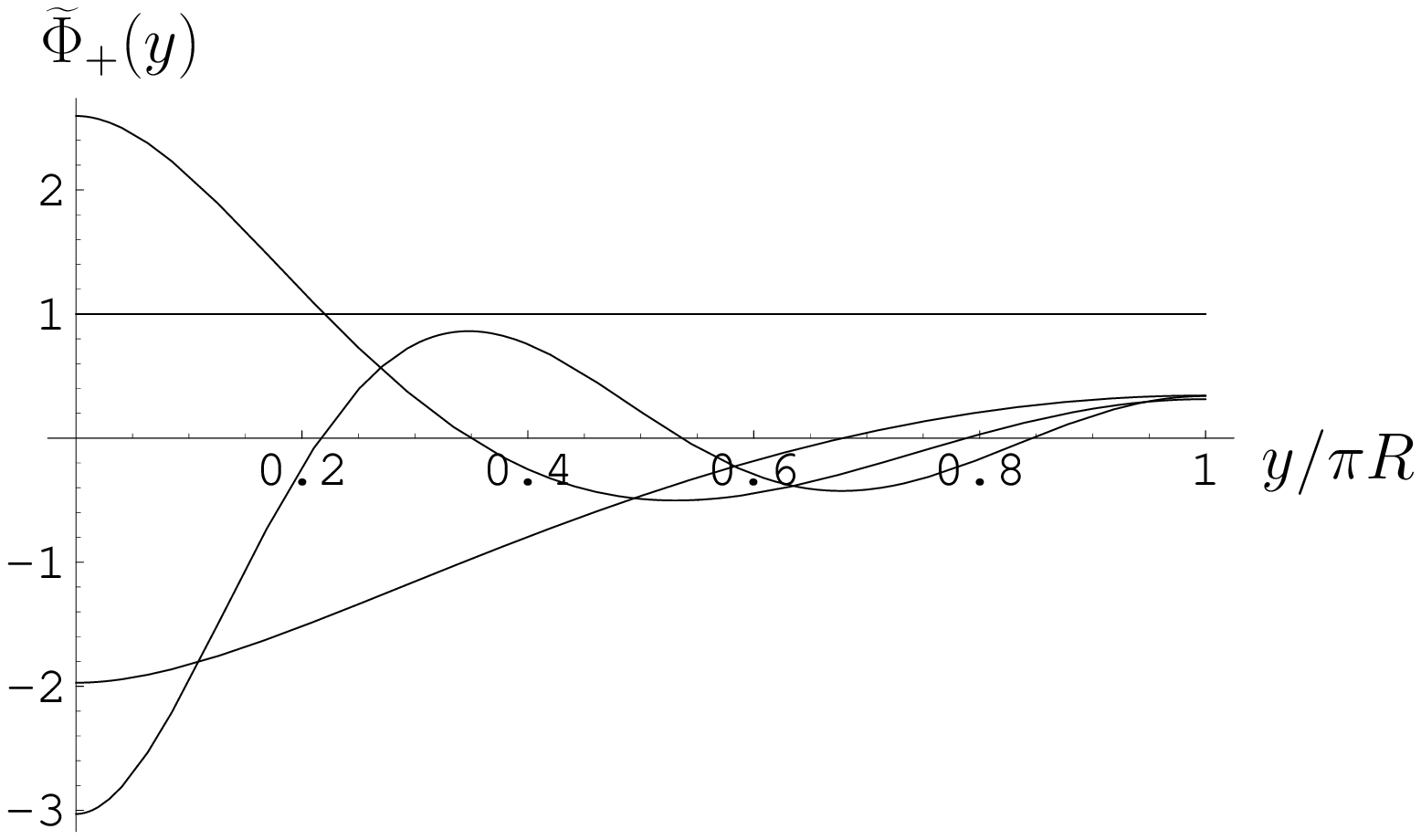,width=\linewidth}}
\end{minipage}
\hfill
\begin{minipage}{0.43\linewidth}
   \centerline{\epsfig{figure=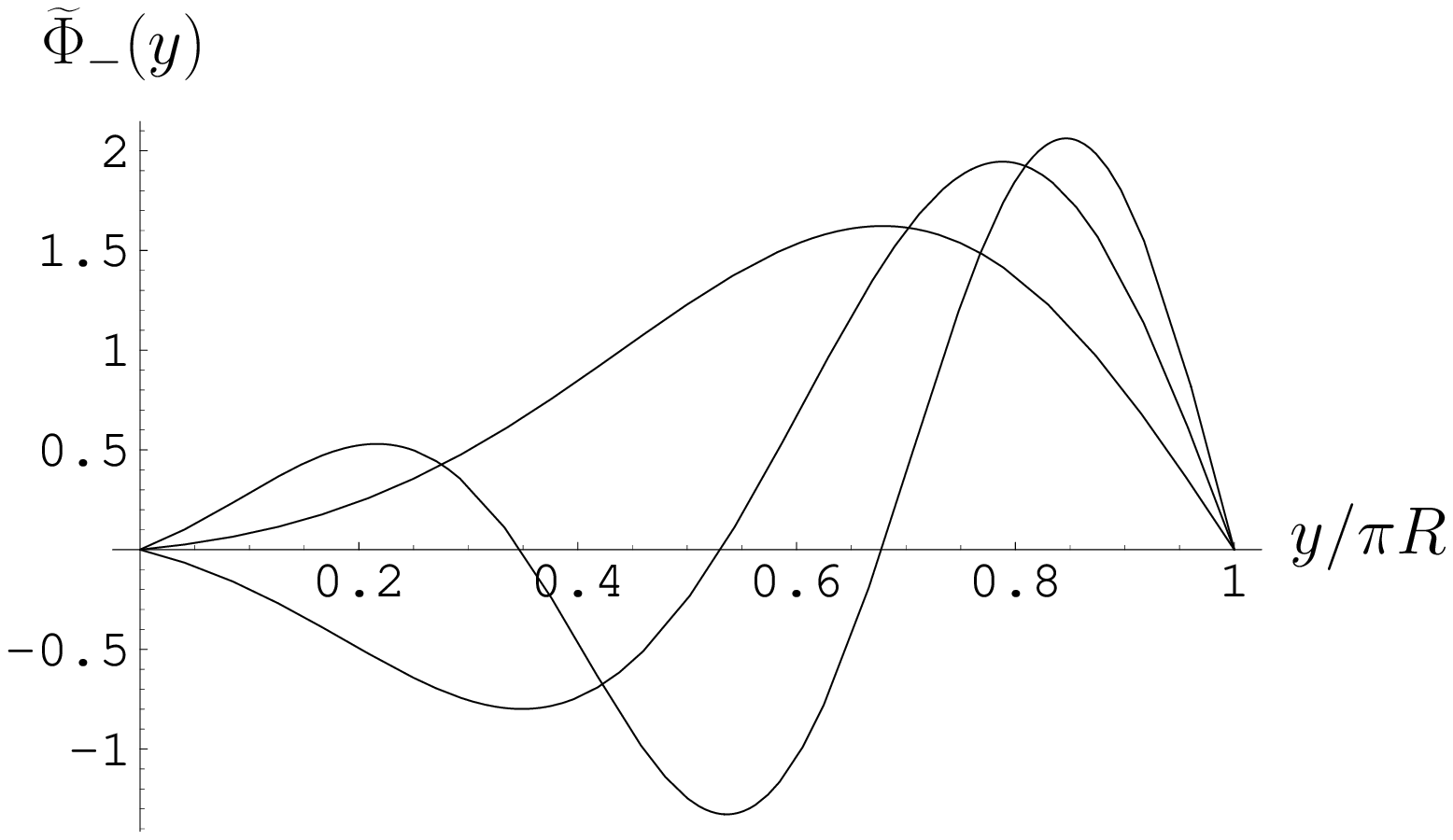,width=\linewidth}}
\end{minipage}
\end{center}
\begin{center}
$\left\{ 
\begin{array}{l}
z_0=5 \\ z_\pi=0 
\end{array}
\right.$
\hfill
\begin{minipage}{0.43\linewidth}
   \centerline{\epsfig{figure=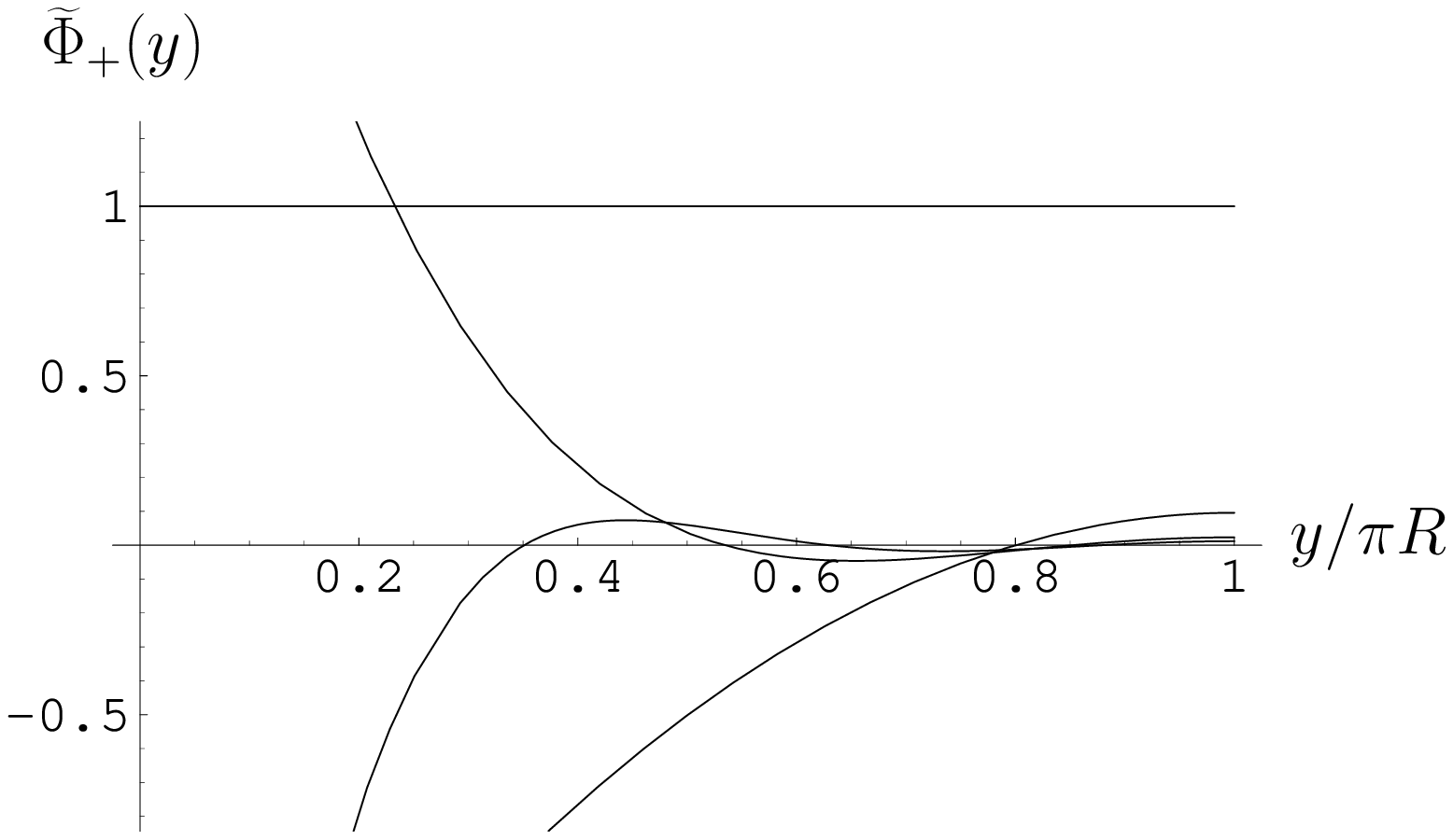,width=\linewidth}}
\end{minipage}
\hfill
\begin{minipage}{0.43\linewidth}
   \centerline{\epsfig{figure=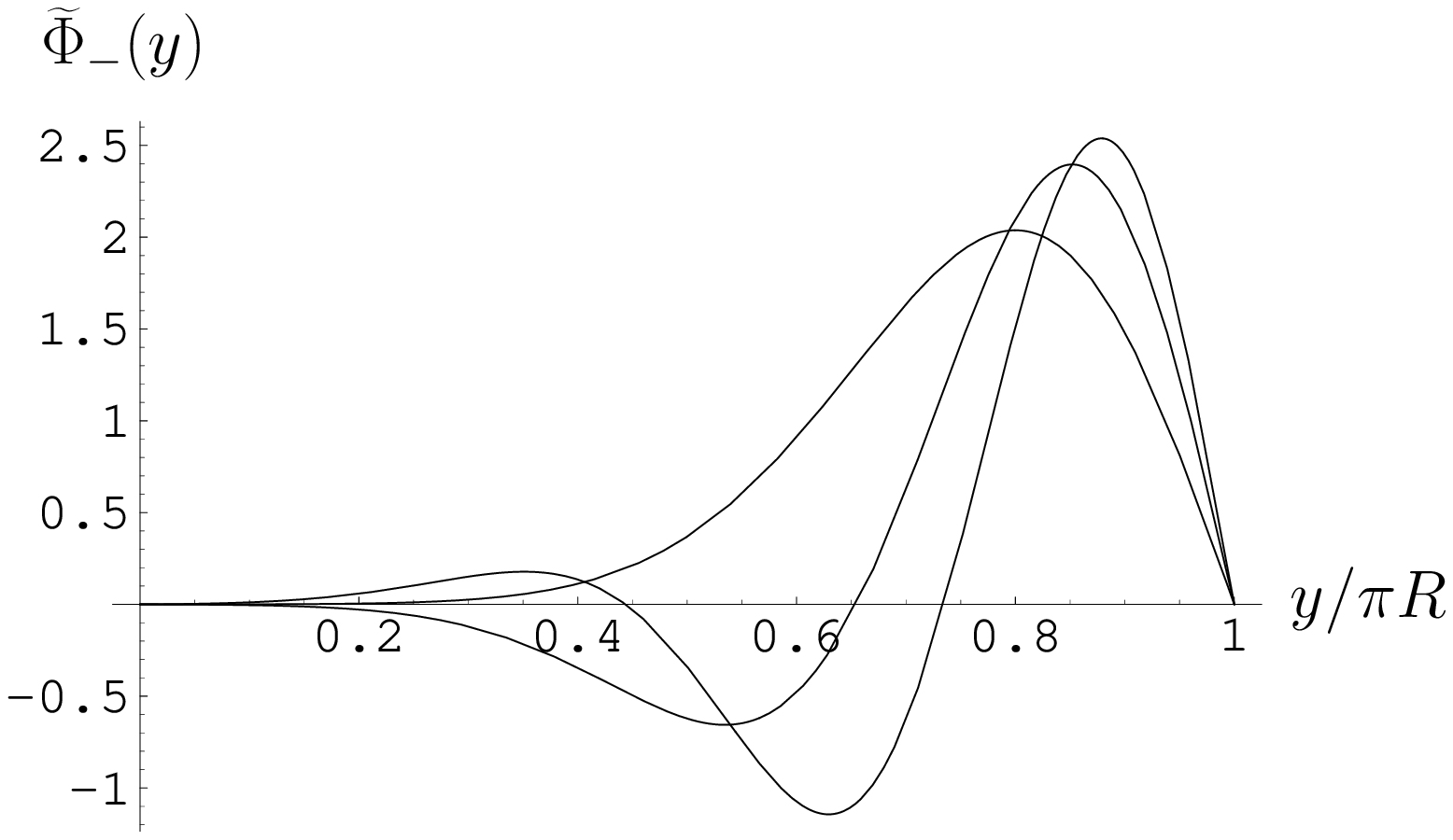,width=\linewidth}}
\end{minipage}
\end{center}
\caption{Wave function profile of $\widetilde\Phi_\pm$ up to 
the 3rd excited mode, with several boundary parameters 
$z_0$ ($z_\pi \equiv 0$) defined by Eq.~(\ref{defz}). 
The zero mode belongs to $\Phi_+$, 
and it is flat with respect to $\widetilde\Phi_+$. 
To obtain the wave function 
profile of the original field, $\Phi_\pm$, all $\widetilde\Phi_\pm (y)$ 
depicted here should be multiplied by the Gaussian factor 
(\ref{profile-3}) (see Eq.~(\ref{redeftp})).}
\label{fig:wf2}
\end{figure}

\begin{figure}[t]
\begin{center}
$\left\{ 
\begin{array}{l}
z_0=1/\sqrt{2} \\ z_\pi=1/\sqrt{2} 
\end{array}
\right.$
\hfill
\begin{minipage}{0.41\linewidth}
   \centerline{\epsfig{figure=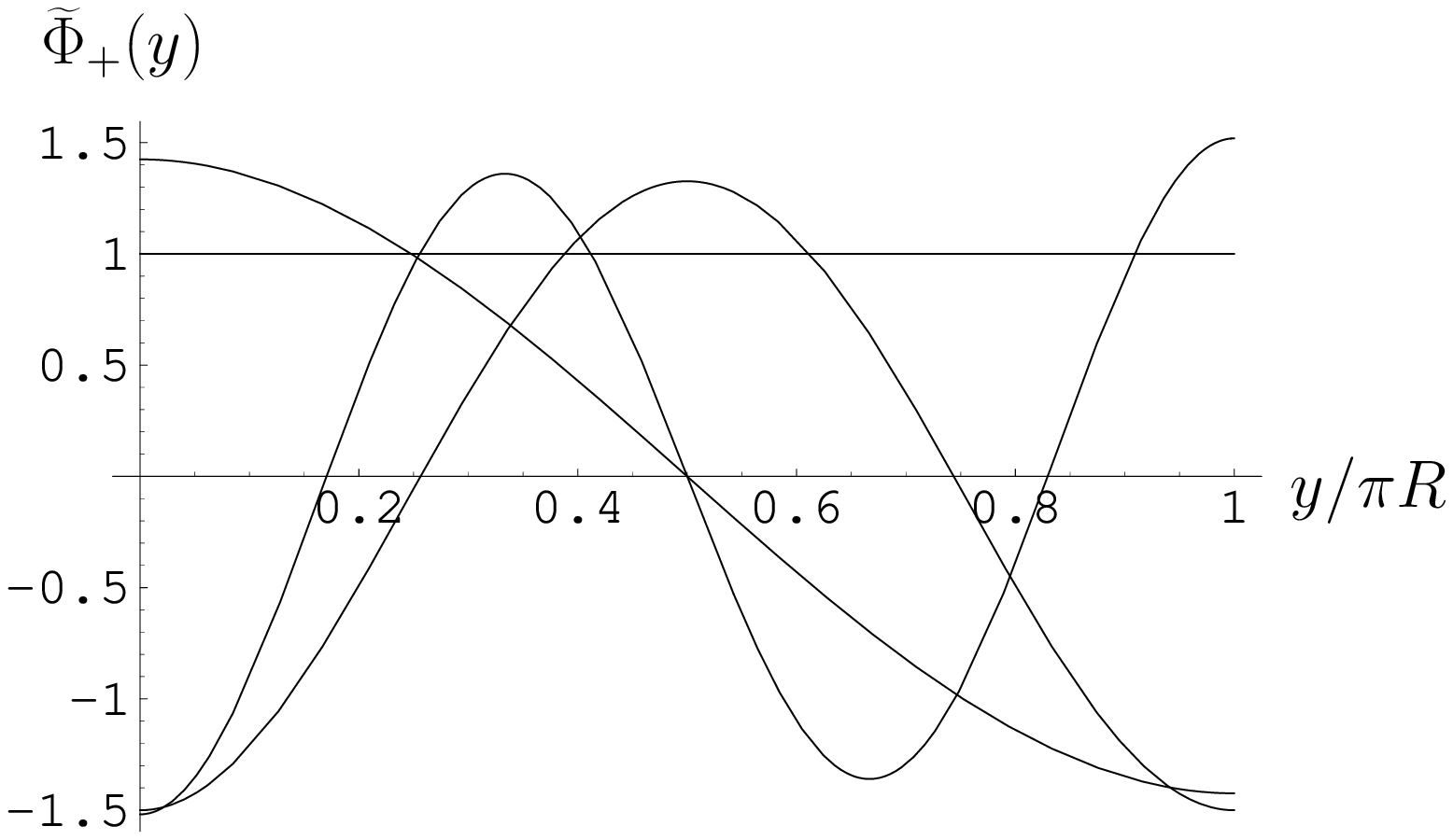,width=\linewidth}}
\end{minipage}
\hfill
\begin{minipage}{0.41\linewidth}
   \centerline{\epsfig{figure=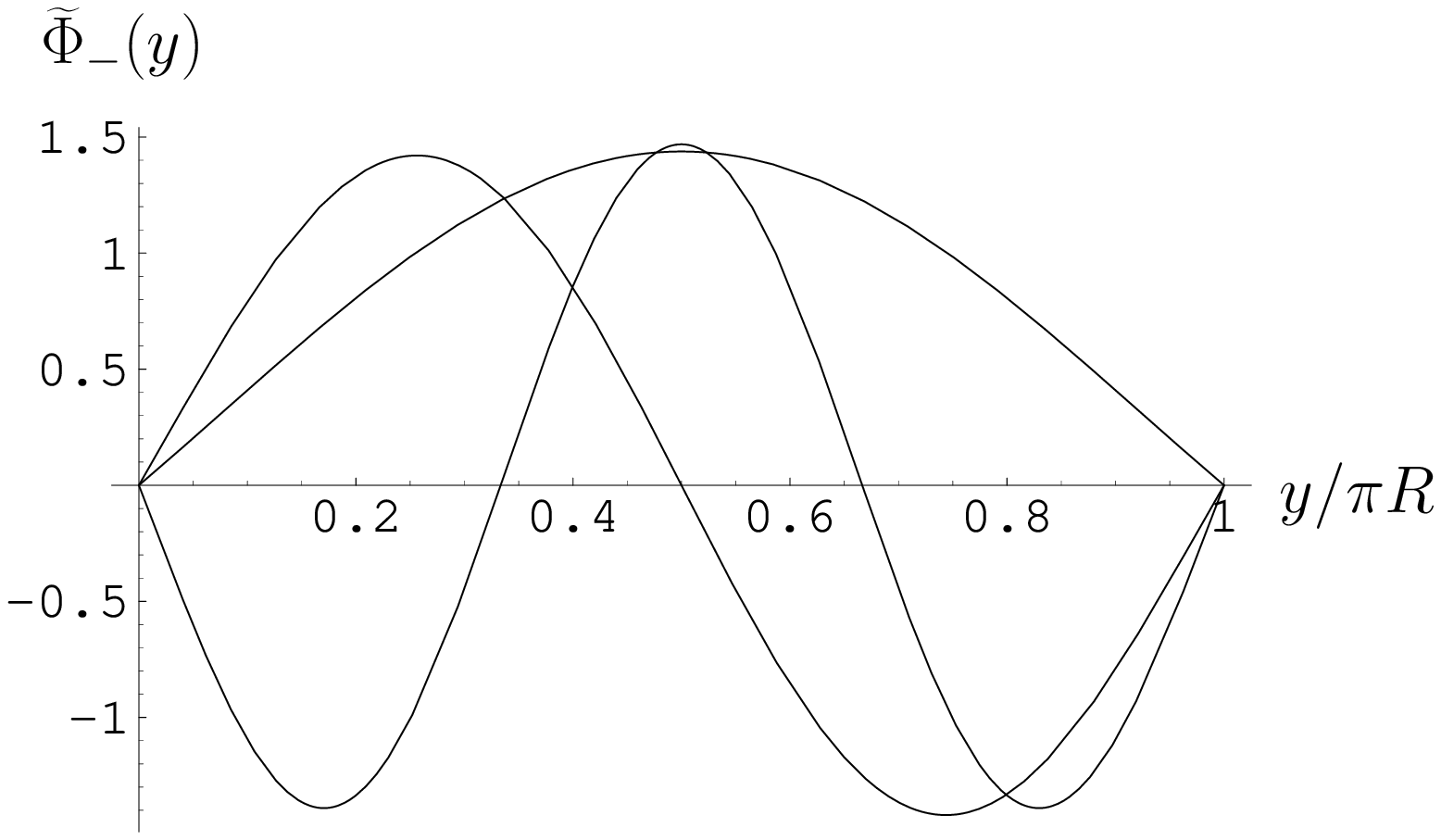,width=\linewidth}}
\end{minipage}
\end{center}
\begin{center}
$\left\{ 
\begin{array}{l}
z_0=3/\sqrt{2} \\ z_\pi=3/\sqrt{2}
\end{array}
\right.$
\hfill
\begin{minipage}{0.41\linewidth}
   \centerline{\epsfig{figure=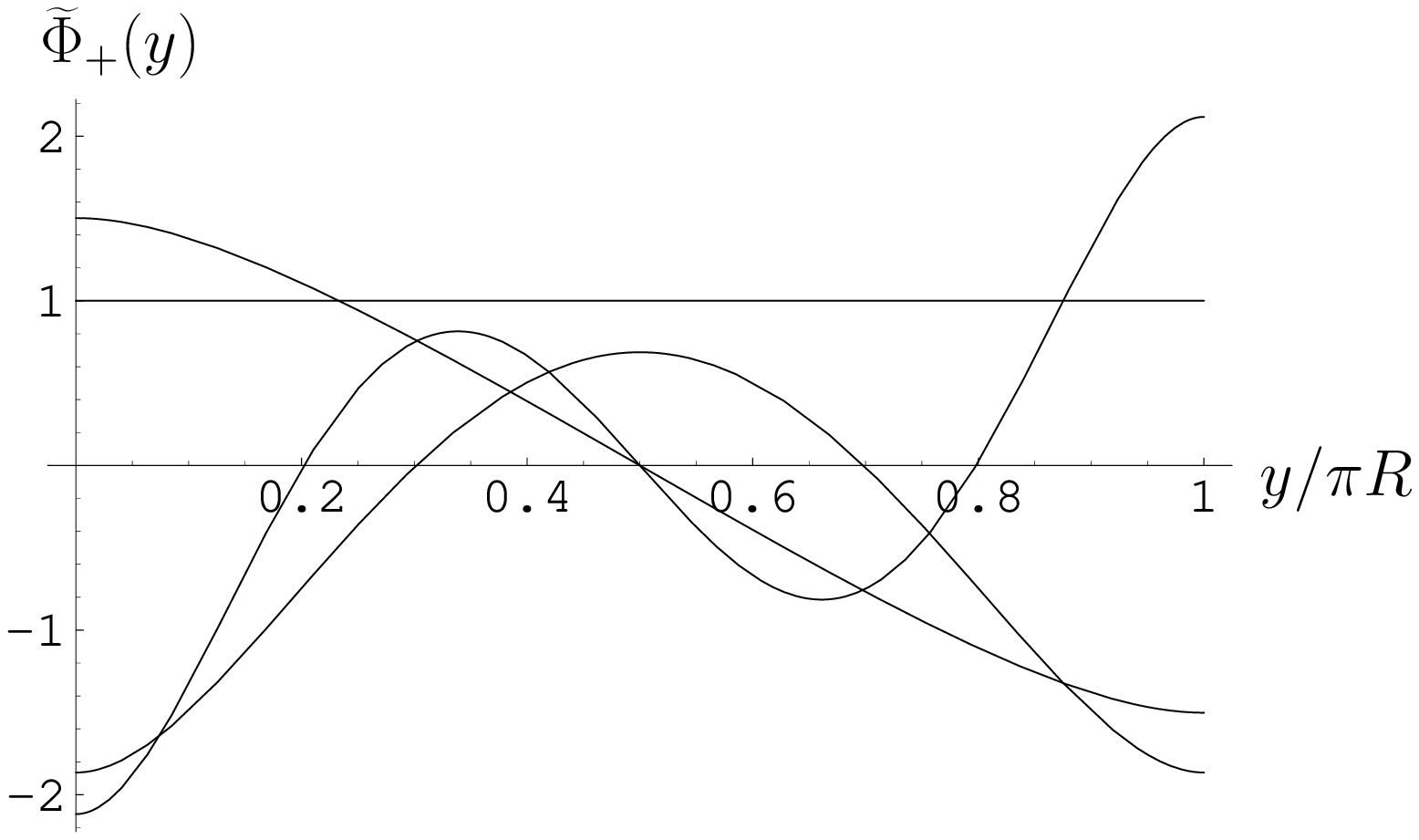,width=\linewidth}}
\end{minipage}
\hfill
\begin{minipage}{0.41\linewidth}
   \centerline{\epsfig{figure=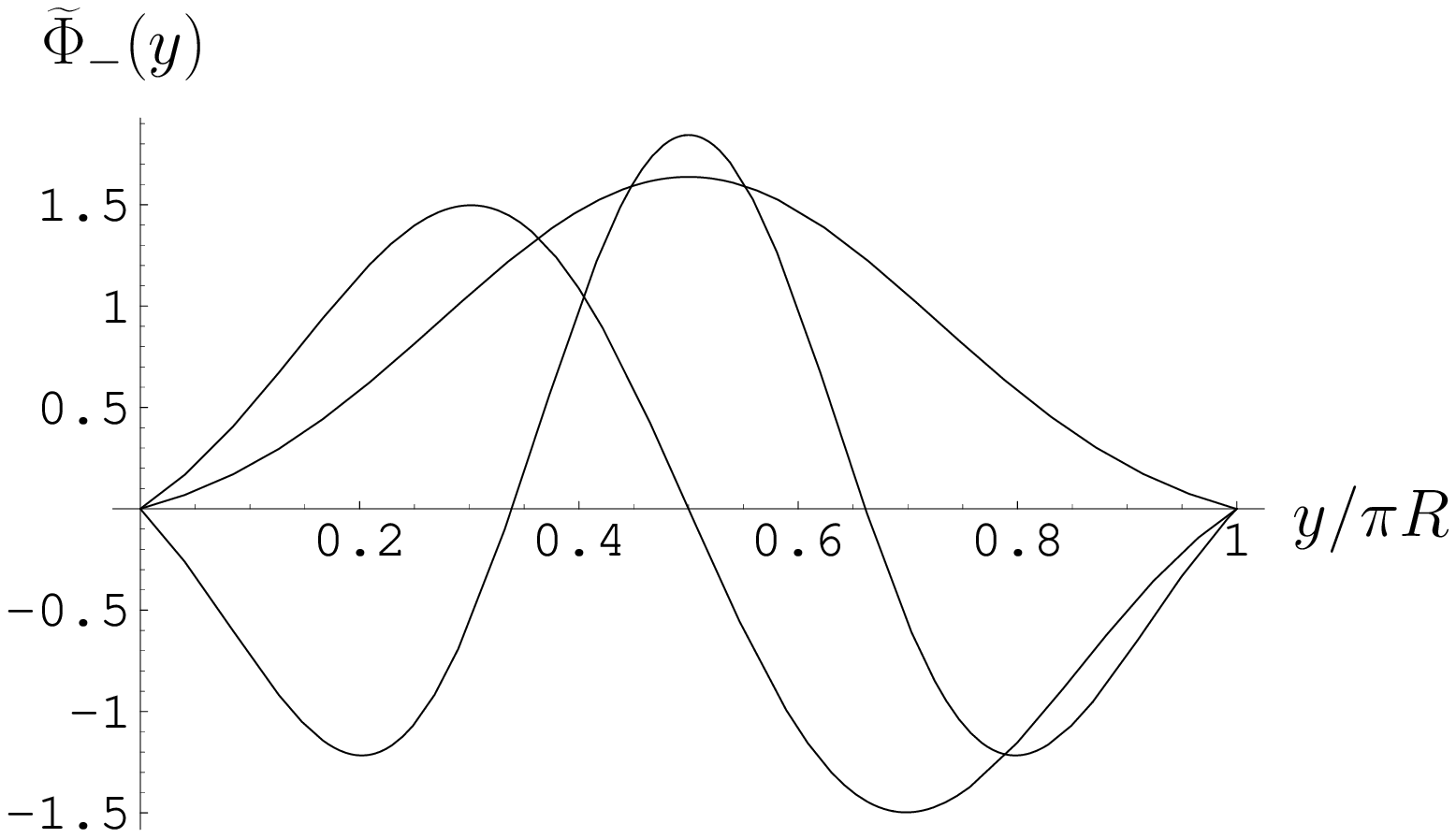,width=\linewidth}}
\end{minipage}
\end{center}
\begin{center}
$\left\{ 
\begin{array}{l}
z_0=5/\sqrt{2} \\ z_\pi=5/\sqrt{2}
\end{array}
\right.$
\hfill
\begin{minipage}{0.41\linewidth}
   \centerline{\epsfig{figure=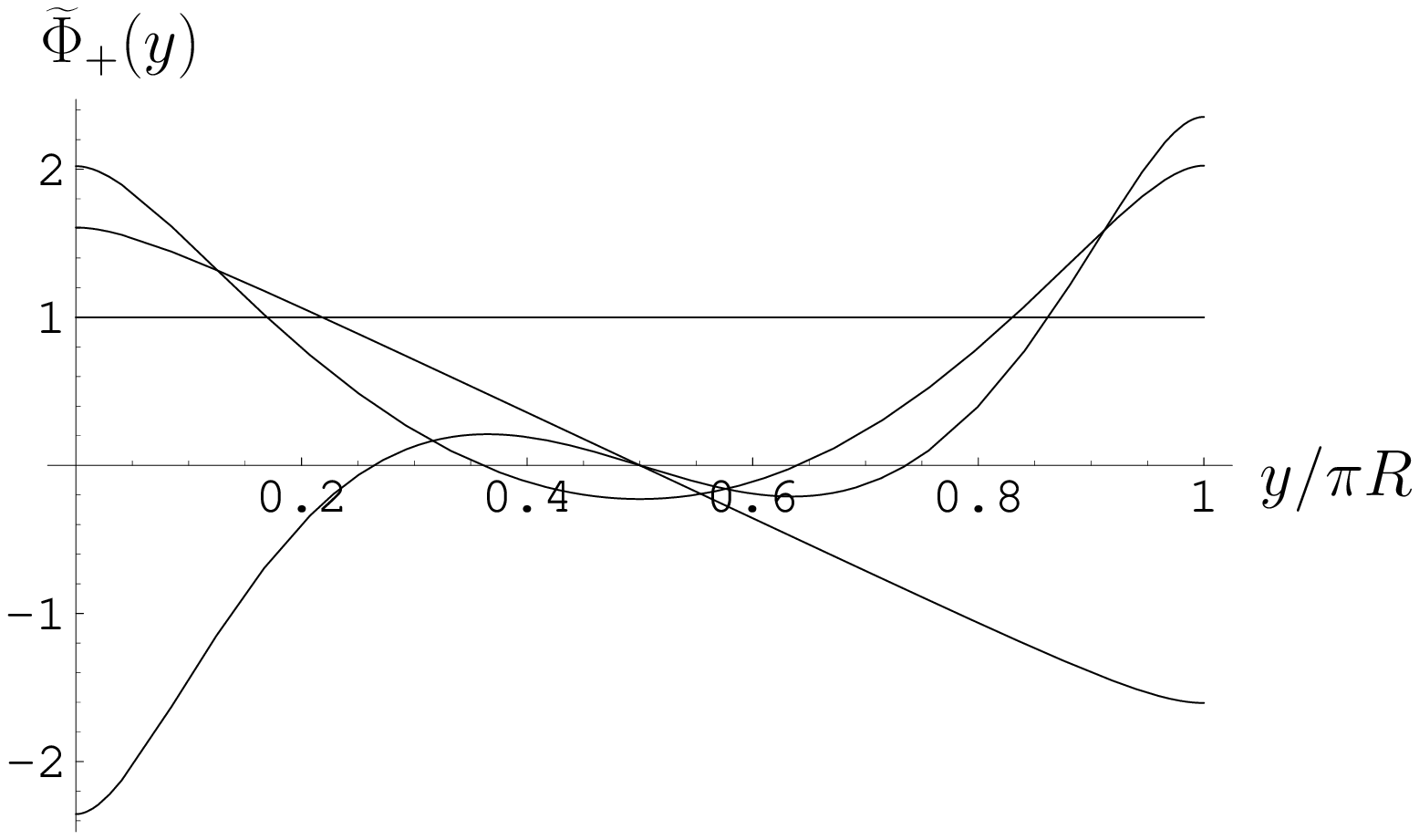,width=\linewidth}}
\end{minipage}
\hfill
\begin{minipage}{0.41\linewidth}
   \centerline{\epsfig{figure=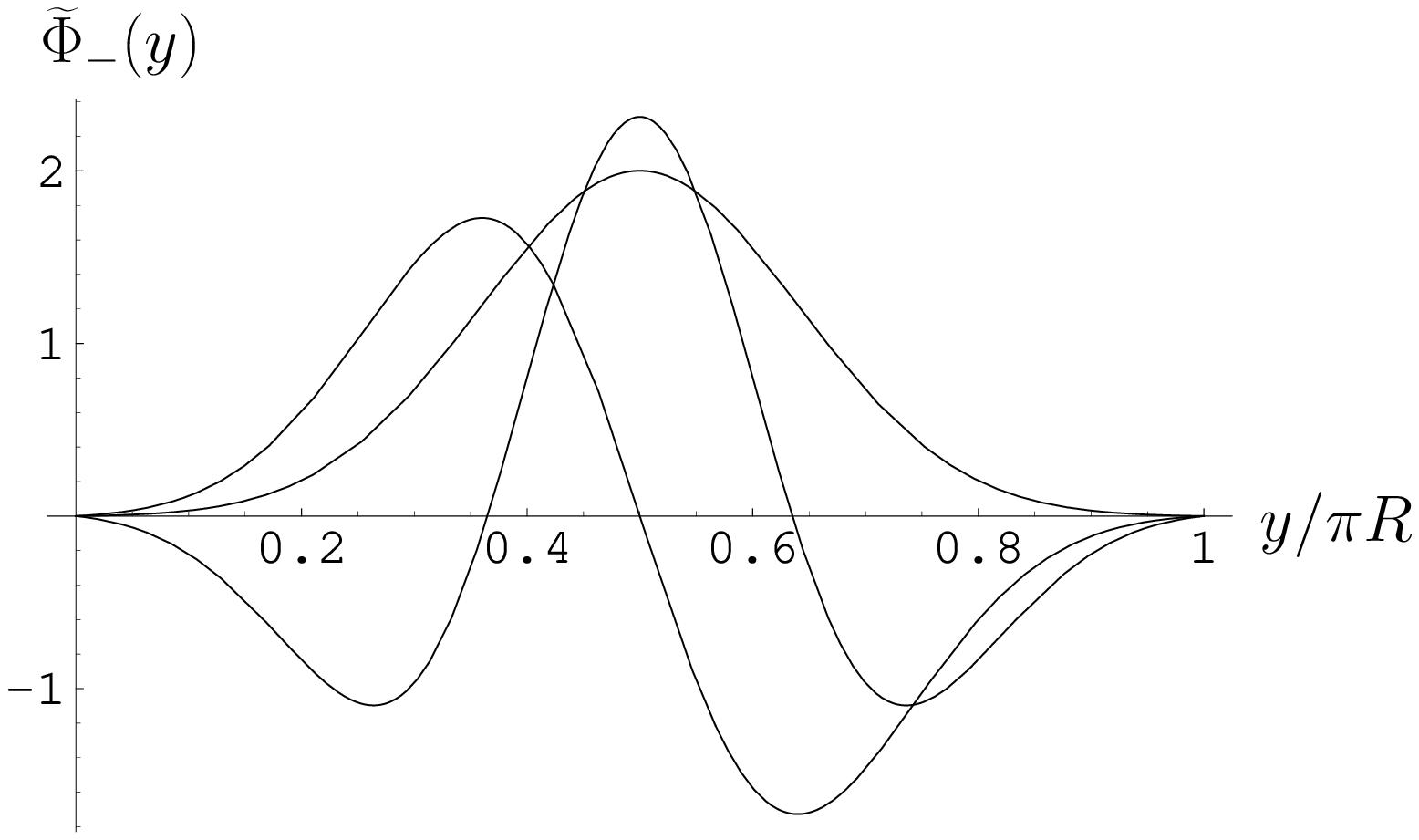,width=\linewidth}}
\end{minipage}
\end{center}
\caption{Wave function profile of $\widetilde\Phi_\pm$ up to 
the 3rd excited mode, with several boundary parameters 
$z_0=z_\pi$ defined by Eq.~(\ref{defz}). 
The zero mode belongs to $\Phi_+$, 
and it is flat with respect to $\widetilde\Phi_+$. 
To obtain the wave function 
profile of the original field, $\Phi_\pm$, all $\widetilde\Phi_\pm (y)$ 
depicted here should be multiplied by the Gaussian factor 
(\ref{profile-3}) (see Eq.~(\ref{redeftp})).}
\label{fig:wf1}
\end{figure}

{}From the spectrum, we can also derive the wave function profile of 
$\widetilde\Phi_\pm$. This is shown in Figs.~\ref{fig:wf2} 
and \ref{fig:wf1}. 
These figures reveal that the higher modes of $\widetilde\Phi_+$ 
($\widetilde\Phi_-$) localize toward (away from) the brane(s) 
with non-zero FI-term(s). 
By fixing all the parameters except for $q$ 
in $z_0$ and $z_\pi$, we also find that the localization 
profile becomes sharp as the charge $q$ increases. 

\subsection*{The case with $\langle \Sigma \rangle, \langle 
\Phi_+ \rangle \neq 0$ ($^\exists q<0$)}

Next, we analyze the case with $q<0$ for only one bulk field $\Phi_+$, 
as studied in \S 2.2.3. 
We consider another bulk field, $\Phi'_\pm$, with charge $\pm q'$ 
in such a configuration. 
{}From the potential (\ref{potential}) we have a $\Phi'_\pm$ square 
term after $\Sigma$ and $\Phi_+$ develop VEVs $\langle \Sigma \rangle$ 
and $\langle \Phi_+ \rangle$. 
It is written 
\begin{eqnarray}
-{\cal L}_{\rm m}^{\Phi'_\pm} &=& 
|\partial_y \Phi'_\pm \mp gq' \langle \Sigma \rangle \Phi'_\pm|^2
\nonumber \\ &=& 
-\Phi_\pm'^\dagger \left( \partial_y^2 
 \mp gq' (\partial_y \langle \Sigma \rangle) 
 - (gq')^2 \langle \Sigma \rangle^2 
\right) \Phi'_\pm 
\equiv 
-\Phi_\pm'^\dagger \Delta_\pm \Phi'_\pm, 
\label{KKop}
\end{eqnarray}
where $\Delta_\pm \equiv \partial_y^2 
 \mp gq' (\partial_y \langle \Sigma \rangle) 
 - (gq')^2 \langle \Sigma \rangle^2$. 
The eigenvalues of the operator 
$\Delta_\pm$ correspond to the KK spectrum of $\Phi'_\pm$. 

We again solve the eigenvalue equation (\ref{PhiEE}). 
The VEVs $\partial_y \langle \Sigma \rangle$ and 
$\langle \Sigma \rangle$ in $\Delta_\pm$ are derived from 
Eqs.~(\ref{D-flat3}), (\ref{vevphi-3}) and (\ref{vevSigma-3}). 
Then, defining $z \equiv \tan (ay+b+c_0)$, in the bulk ($0<y<\pi R$), 
the eigenvalue equation becomes 
\begin{eqnarray}
(1+z^2) \partial_z^2 \Phi'_\pm (z) 
+ 2z \partial_z \Phi'_\pm (z) 
-  \left( n_\pm(n_\pm+1) - \frac{\nu_\pm}{1+z^2} \right) 
\Phi'_\pm (z) =0, 
\end{eqnarray}
where $n_\pm(n_\pm+1)=q'(q'\pm q)/q^2$ and 
$\nu_\pm=\lambda/a^2 \pm q'/q +n_\pm(n_\pm+1)$. 
Thus we obtain 
\begin{eqnarray}
\Phi'_\pm (z) = 
  A_\pm P_{n_\pm}^{\nu_\pm} (iz) 
+ B_\pm Q_{n_\pm}^{\nu_\pm} (iz),
\end{eqnarray}
where $P_n^\nu (x)$ ($Q_n^\nu (x)$) is the first (second) class 
associated Legendre function, and $A_\pm$ and $B_\pm$ are constants. 
The mass eigenvalue $\lambda$ can be obtained 
by requiring that $A_\pm$ and $B_\pm$ cannot be zero 
simultaneously in the boundary conditions, as in the previous case.


\end{document}